\documentclass[11pt,english,vecarrow]{extarticle}
\usepackage[T1]{fontenc}
\usepackage[utf8]{inputenc}
\usepackage[letterpaper]{geometry}
\usepackage{babel}
\usepackage{float}
\usepackage{amsmath}
\usepackage{amssymb}
\usepackage{graphicx}
\usepackage{setspace}

\makeatletter
\numberwithin{equation}{section}
\numberwithin{figure}{section}
\numberwithin{table}{section}

\usepackage[titletoc]{appendix}
\usepackage{bbm}
\usepackage{tensind}
\tensordelimiter{?}
\usepackage{slashed}
\usepackage{multicol}
\usepackage[labelfont={sc,small},textfont=small]{caption}
\newcommand{\overbar}[1]{\mkern 1.5mu\overline{\mkern-1.5mu#1\mkern-1.5mu}\mkern 1.5mu}

\makeatletter
\newcommand{\qbar}{\text{\q@bar}}
\newcommand{\q@bar}{%
  \vphantom{$\m@th q$}%
  \ooalign{%
    $\m@th q$\cr
    \hidewidth\smash{\raisebox{-1.55ex}{$\m@th\mathchar'26$}}\hidewidth\cr}%
}
\makeatother
\makeatletter
\newcommand{\lambdabar}{\text{\lambda@bar}}
\newcommand{\lambda@bar}{%
  \vphantom{$\m@th \lambda$}%
  \ooalign{%
    $\m@th \lambda$\cr
    \hidewidth\smash{\raisebox{-0.05ex}{$\m@th\mathchar'26$}}\hidewidth\cr}%
}
\makeatother

\makeatother

\begin{document}
\title{Stress energy momentum in terms of geodesic accelerations and variational
tensors including torsion\\
\quad{}}
\author{\doublespacing{}\textsc{Adam Marsh}\thanks{\textsc{Electronic address: adammarsh@berkeley.edu}}}
\date{\ }
\maketitle
\begin{abstract}
General relativity and its extensions including torsion identify stress
energy momentum as being proportional to the Einstein tensor, thus
ensuring both symmetry and conservation. Here we visualize stress
energy and momentum by identifying the associated relative fractional
accelerations of geodesics encoded in the Einstein tensor. This also
provides an intuitive explanation for the vanishing divergence of
the Einstein tensor. In order to obtain this same energy and momentum
for other actions such as that of Dirac theory including torsion,
we then review the various stress energy momentum tensors resulting
from the variation of different quantities derived from parallel transport,
and detail their interrelationships. This provides an opportunity
to revisit some classic material from a geometric point of view, including
Einstein-Cartan theory, the Sciama-Kibble formalism, and the Belinfante-Rosenfeld
relation, whose derivation in the mostly pluses signature would seem
to not be otherwise readily available. 
\end{abstract}
\tableofcontents{}

\section{Introduction}

\subsection{Motivation}

The components of the Hilbert stress energy momentum (SEM) tensor
are commonly named using terms such as ``energy density'' and ``momentum
density'' which are derived from continuum mechanics; but these intuitive
meanings derived from mechanical systems are not applicable to other
theories such as electromagnetism. Since the gravitational equations
of motion (EOM) specify the Hilbert SEM tensor as being proportional
to the (torsionless) Einstein tensor, we may instead associate these
terms with the fractional accelerations of geodesics described by
the components of the Einstein tensor. Here we detail these associations,
our motivation being that unlike those derived from continuum mechanics,
they survive the transition to electromagnetism and the theories beyond
it.

We would then like to determine these geodesic acceleration defined
forms of energy and momentum for theories whose matter actions may
depend on the spacetime connection in ways other than via the associated
metric; we would also like to consider connections which may have
non-zero torsion. We provide a general treatment, but the obvious
target action for this analysis is that of Dirac theory. We anticipate
two main benefits from such a treatment: first, a set of results which
may be applied to Dirac theory in subsequent work; and second, a modern,
detailed, and consistent treatment of classic material which, especially
from a geometric viewpoint in the mostly pluses signature, would seem
to not be otherwise readily available (the closest reference we find
from an algebraic viewpoint in the mostly minuses signature is \cite{Poplawski}). 

\subsection{Overview}

In Section \ref{sec:Energy-momentum-as-geodesic-acceleration} we
detail and visualize energy, momentum, and stress as defined by the
components of the Einstein tensor in terms of the fractional accelerations
of geodesics. After treating energy and momentum, we use their geometric
meanings in Section \ref{subsec:energy-conservation-and-divG-vanishing}
to arrive at an intuitive explanation for the vanishing divergence
of the Einstein tensor in terms of local conservation of energy. In
our subsequent treatment of pressure in Section \ref{subsec:Pressure},
we also note the odd asymmetry between time-like and space-like geodesic
accelerations in the presence of mass, and contrast it with other
forms of matter for which this asymmetry is absent. 

In Section \ref{sec:Geometry} we review coordinate and orthonormal
frames and their connections, and then treat Lorentz indices and various
related expressions in some detail from a geometric point of view.
We also consider the variations of the different quantities derived
from spacetime parallel transport, along with their interdependencies,
an analysis which would seem to be difficult to find in the literature.

In Section \ref{sec:SEM-tensors} we then define various SEM tensors
and their interrelationships; inspired by \cite{GockelerSchucker},
we work in terms of differential forms when treating Einstein-Cartan-Sciama-Kibble
theory in Section \ref{subsec:The-tetrad-SEM-tensor}. Our end goal
is the Belinfante-Rosenfeld relation, derived in Section \ref{subsec:The-Belinfante-Rosenfeld-relation}
with an alternate derivation in Appendix \ref{sec:alt-Belinfante-Rosenfeld-derivation},
which allows us to obtain the SEM tensor proportional to the torsionless
Einstein tensor from the SEM tensors obtained when varying the frame
and spin connection in theories such as Dirac theory. Section \ref{sec:Summary}
summarizes these results.

Throughout the paper we will use the mostly pluses spacetime metric
signature, where in an orthonormal frame the metric is $g_{\mu\nu}=\mathrm{diag}\left(-1,1,1,1\right)$.
We will strive to distinguish quantities related to torsionless parallel
transport with an overbar (e.g. $\overline{G}$ for the torsionless
Einstein tensor), and to standardize our index notation according
to the following:
\begin{itemize}
\item Spacetime indices: $\kappa,\lambda,\mu,\nu,\rho,\sigma$ 
\item Space indices: $\mathsf{i},\mathsf{j},\mathsf{k},\mathsf{l},\mathsf{m},\mathsf{n}$ 
\item Lorentz indices: $i,j,k,l,m,n,p,q,r,s$ 
\end{itemize}
We also adopt the notation from \cite{Marsh-book} in which an arrow
decoration, e.g. $\vec{\Phi}$, indicates a vector- or $\mathbb{K}^{n}$-valued
form, where $\mathbb{K}$ is either $\mathbb{R}$ or $\mathbb{C}$,
while a check decoration, e.g. $\check{\Gamma}$ indicates an algebra-
or matrix-valued form; in Section \ref{subsec:The-tetrad-SEM-tensor}
only, we will also denote forms with an underbar in order to make
clear the types of each quantity being considered. Finally, we will
bold key terms when first defined.

\section{\label{sec:Energy-momentum-as-geodesic-acceleration}Energy momentum
as geodesic acceleration}

In general relativity, the spacetime manifold is assumed to include
a pseudo-Riemannian metric, and the EOM obtained by varying this metric
in the action,
\begin{equation}
\overline{G}^{\mu\nu}=\kappa\overline{T}^{\mu\nu},
\end{equation}
where in SI units
\begin{equation}
\kappa\equiv\frac{8\pi G}{c^{4}},
\end{equation}
requires the Einstein tensor to be proportional to the Hilbert SEM
tensor of matter. The components of the Hilbert SEM tensor take their
names via identifications with energy and momentum densities and currents
based upon the physical model of continuum mechanics. In electromagnetism,
however, the Hilbert SEM tensor has no such obvious physical interpretation,
but the proportionality of its components to those of $\overline{G}$
remains, and therefore so does their identification with the acceleration
of geodesics described by these components. Hence, if we continue
to describe the Hilbert SEM tensor components in terms of energy and
momentum (as is common), then we are in effect \textit{defining} energy
and momentum by the acceleration of geodesics near matter (including
mass and electromagnetic fields). This idea is what we pursue in this
section. 

\subsection{The Einstein tensor }

In this section all tensors are those corresponding to a connection
with zero torsion, and to avoid clutter we therefore drop the overbar
decoration used in the remainder of the paper to indicate this. 

Recall (see e.g. \cite{Marsh-book} Section 9.3.5) that in an orthonormal
frame (tetrad) $e_{i}$ we have

\begin{equation}
\begin{aligned}G_{ji}=G_{ij} & =G\left(e_{i},e_{j}\right)\\
 & =R^{k}{}_{ikj}-\frac{R}{2}\eta_{ij}\\
 & =\mathrm{Ric}\left(e_{i},e_{j}\right)-\frac{R}{2}\eta_{ij}\\
 & =\sum_{\begin{subarray}{c}
k\neq i\\
k\neq j
\end{subarray}}\eta_{kk}\left\langle \check{R}(e_{k},e_{j})\vec{e}_{i},e_{k}\right\rangle -\eta_{ij}\sum_{\begin{subarray}{c}
m<n\\
\\
\end{subarray}}\eta_{mm}\eta_{nn}\left\langle \check{R}(e_{m},e_{n})\vec{e}_{n},e_{m}\right\rangle \\
\Rightarrow G_{ii} & =-\eta_{ii}\sum_{\begin{subarray}{c}
m<n\\
m,n\neq i
\end{subarray}}\eta_{mm}\eta_{nn}\left\langle \check{R}(e_{m},e_{n})\vec{e}_{n},e_{m}\right\rangle .
\end{aligned}
\end{equation}
Here $G$ is the Einstein tensor, $\mathrm{Ric}$ is the Ricci tensor,
$R$ is the scalar curvature, and $\check{R}$ is the $so(3,1)$-valued
curvature 2-form. Under our zero torsion assumption, $\mathrm{Ric}$
is symmetric and thus so is $G$.

The quantity
\begin{equation}
\begin{aligned}\eta_{mm}\eta_{nn}\left\langle \check{R}(e_{m},e_{n})\vec{e}_{n},e_{m}\right\rangle  & =\eta_{mm}\eta_{nn}R_{mnmn}\end{aligned}
\end{equation}
is the sectional curvature, which depends only upon the plane defined
by $e_{m}$ and $e_{n}$. It is symmetric in the two indices and vanishes
when they are equal. Geometrically, it is the acceleration of two
parallel geodesics in the $e_{n}$ direction with initial infinitesimal
separation $L$ in the direction $e_{m}$ towards each other as a
fraction of $L$ (see \cite{Marsh-book} Section 9.3.6): 
\begin{equation}
\begin{aligned}\eta_{mm}\eta_{nn}\left\langle \check{R}(e_{m},e_{n})\vec{e}_{n},e_{m}\right\rangle  & =\frac{\partial_{n}^{2}L}{L}.\end{aligned}
\end{equation}
The use of the term ``acceleration'' used here thus does not imply
that $e_{n}$ is a time-like direction. This geometric interpretation
is what allows us to characterize the Einstein tensor components in
terms of fractional geodesic accelerations. 

\subsection{Energy}

\begin{figure}[H]
\noindent \begin{centering}
\includegraphics{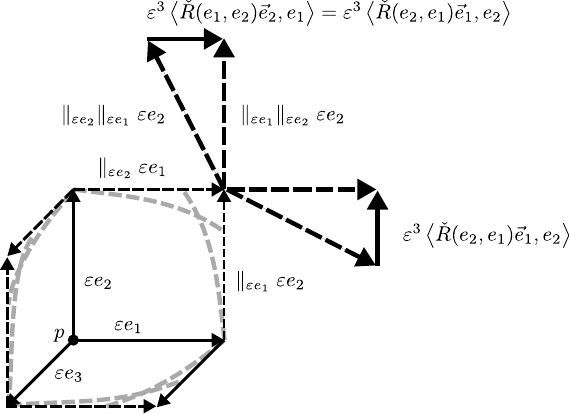}
\par\end{centering}
\caption{\label{fig:gr-energy}A positive energy density $T^{00}\propto G_{00}$
at $p$ is the sum of the fractional accelerations of the space-like
geodesics surrounding $p$ (depicted as dashed grey curves) towards
each other. As shown, the acceleration of the orthogonal geodesic
in the same plane is equal due to the symmetry of the sectional curvature.
Note that the sectional curvatures also remain constant under the
change of sign of any of the argument vectors, resulting in the symmetry
of the next figure.}
\end{figure}

\begin{figure}[H]
\noindent \begin{centering}
\includegraphics{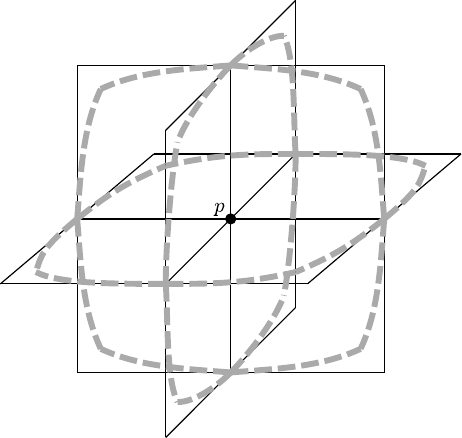}
\par\end{centering}
\caption{\label{fig:energy-density-3d}A positive energy density $T^{00}$
at $p$ causes \textquotedblleft straight\textquotedblright{} objects
(space-like geodesics) to \textquotedblleft bend around\textquotedblright{}
$p$ (if the curvature is isotropic as assumed here). So as compared
to an infinitesimal cube around $p$ constructed using parallel transported
frame components, an infinitesimal cube constructed using geodesics
has faces which curve inwards, reducing its volume and surface area.}
\end{figure}

\begin{figure}[H]
\noindent \begin{centering}
\includegraphics{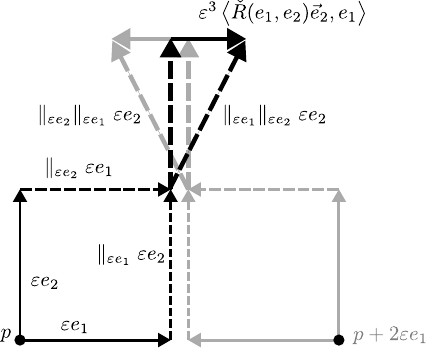}
\par\end{centering}
\caption{\label{fig:energy-adjacent}For a uniform energy density, one may
ask how the geodesics can \textquotedblleft bend around\textquotedblright{}
adjacent points at the same time. The answer lies in remembering that
the \textquotedblleft bending\textquotedblright{} is relative to the
cube we construct out of parallel transported vectors; if we construct
an adjacent such cube, the overlapping parallel transported vectors
will not coincide, resulting in a different depiction of parallel
cube edges in the figure such that the geodesic \textquotedblleft bends\textquotedblright{}
towards each point relative to its cube edge.}
\end{figure}

The energy density is
\begin{align}
T^{00}\propto G_{00} & =\sum_{\begin{subarray}{c}
m<n\\
m,n\neq i
\end{subarray}}\left\langle \check{R}(e_{m},e_{n})\vec{e}_{n},e_{m}\right\rangle \\
 & =\left\langle \check{R}(e_{1},e_{2})\vec{e}_{2},e_{1}\right\rangle +\left\langle \check{R}(e_{1},e_{3})\vec{e}_{3},e_{1}\right\rangle +\left\langle \check{R}(e_{2},e_{3})\vec{e}_{3},e_{2}\right\rangle ,
\end{align}
which is also the scalar curvature in the space-like hyperplane defined
by the tetrad. Geometrically, the energy density at a point $p$ is
thus proportional to the total fractional acceleration of the space-like
geodesics surrounding $p$ towards each other (see Figure \ref{fig:gr-energy}).
Qualitatively, positive energy density at $p$ causes ``straight''
extended objects to ``bend around $p$'' (see Figure \ref{fig:energy-density-3d}).
One may wonder how geodesics can ``bend around'' adjacent points
at the same time; this is addressed in Figure \ref{fig:energy-adjacent}.

Note that positive energy density at $p$ does not necessarily mean
time-like geodesics will accelerate towards $p$ (see Section \ref{subsec:Pressure}),
which is our Newtonian expectation for the gravitational force. We
also do not know the amount by which extended objects in different
planes ``bend around'' $p$; we only know the total. In particular,
even with positive energy density, a ``straight'' object may ``bend
away'' from $p$ in a specific plane, as long as the other two planes
``bend around'' $p$ to a greater degree.

\subsection{Momentum}

\begin{figure}[H]
\noindent \begin{centering}
\includegraphics{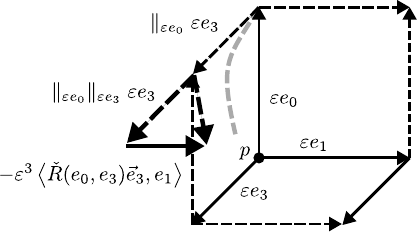}
\par\end{centering}
\caption{\label{fig:gr-momentum}A positive $e_{1}$ momentum density $T^{10}\propto G^{10}$
is the sum of the amounts that future orthogonal space-like geodesics
\textquotedblleft bend towards\textquotedblright{} the $e_{1}$ direction,
as compared to present geodesics parallel transported in the $e_{0}$
direction. Note that unlike with sectional curvature, this quantity
reverses when reversing the sign of $e_{0}$, i.e. past orthogonal
space-like geodesics \textquotedblleft bend away\textquotedblright{}
from the $e_{1}$ direction. }
\end{figure}

\begin{figure}[H]
\noindent \begin{centering}
\includegraphics{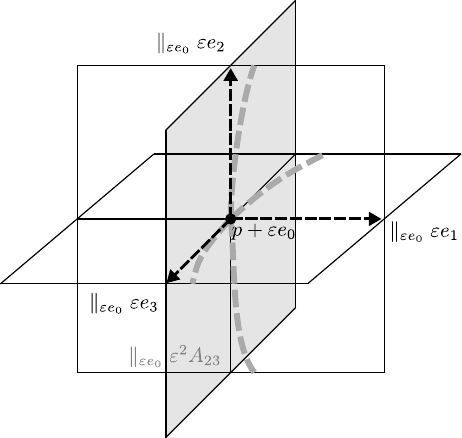}
\par\end{centering}
\caption{\label{fig:momentum-3d}A positive $e_{1}$ momentum density $T^{10}$
at $p$ causes orthogonal \textquotedblleft straight\textquotedblright{}
objects (space-like geodesics) to \textquotedblleft bend towards\textquotedblright{}
the $e_{1}$ direction in the future (if the curvature is isotropic
as assumed here). Note that this \textquotedblleft bending\textquotedblright{}
is the fractional acceleration of future space-like geodesics (depicted
as dashed grey curves) away from present space-like geodesics parallel
transported in the future $e_{0}$ direction. This \textquotedblleft bending\textquotedblright{}
may be viewed as the left edge of the cube in Figure \ref{fig:energy-density-3d},
indicating that in the future there is additional energy density in
the $e_{1}$ direction, i.e. there is a positive energy flux across
$A_{23}$.}
\end{figure}

Using the symmetries of the Riemann curvature tensor for a torsionless
connection (see e.g. \cite{Marsh-book} Section 9.3.2), we see that
the $e_{1}$ component of momentum density is

\begin{equation}
\begin{aligned}T^{10}\propto G^{10} & =\eta^{00}\eta^{11}\sum_{\begin{subarray}{c}
k\neq0\\
k\neq1
\end{subarray}}\eta_{ii}\left\langle \check{R}(e_{k},e_{1})\vec{e}_{0},e_{k}\right\rangle \\
 & =-\sum_{\begin{subarray}{c}
k\neq0\\
k\neq1
\end{subarray}}\left\langle \check{R}(e_{0},e_{k})\vec{e}_{k},e_{1}\right\rangle \\
 & =-\left\langle \check{R}(e_{0},e_{2})\vec{e}_{2},e_{1}\right\rangle -\left\langle \check{R}(e_{0},e_{3})\vec{e}_{3},e_{1}\right\rangle .
\end{aligned}
\label{eq:momentum-density-expression}
\end{equation}
In keeping with the continuum mechanics origin of energy momentum
terminology and the symmetry of the Einstein tensor, the $e_{1}$
momentum density $T^{10}=T^{01}$ is also called the energy current
in the $e_{1}$ direction, or the flux of energy across the surface
$A_{23}\equiv e_{2}\land e_{3}$ orthogonal to $e_{1}$. Geometrically,
the $e_{1}$ momentum density is the sum of the fractional accelerations
of future space-like geodesics in $A_{23}$ towards $e_{1}$ (see
Figure \ref{fig:gr-momentum}). Qualitatively, positive $e_{1}$
momentum density means that future extended objects in the cube face
$A_{23}$ ``bend towards'' the $e_{1}$ direction (see Figure \ref{fig:momentum-3d}).
We can view this as indicating that in the future, there will be additional
energy density present in the $e_{1}$ direction of energy flux, causing
space-like geodesics to ``bend around'' it. Similarly, we can view
the past space-like geodesics ``bending away'' from the $e_{1}$
direction as indicating that in the past, there was additional energy
density present in the $-e_{1}$ direction, causing space-like geodesics
to ``bend around'' it. 

Note that despite the ``bending'' of future geodesics in the $e_{1}$
direction, positive $e_{1}$ momentum density at $p$ does not necessarily
mean future geodesics in the $A_{12}$ and $A_{13}$ planes around
the point $p+\varepsilon e_{1}$ accelerate more towards each other
(thus increasing energy density in the $e_{1}$ direction); this depends
upon whether the geodesics infinitesimally in the $e_{1}$ direction
change their acceleration (which corresponds to the flux of energy
density across $A_{23}$ parallel transported in the $e_{1}$ direction).
In a continuum mechanics, this corresponds to positive $e_{1}$ momentum
density only meaning that infinitesimal particles are moving in the
$e_{1}$ direction, not that there is a greater density of particles
in this direction, since the $e_{1}$ momentum there may be even larger.
We also do not know how much future extended objects in the $e_{2}$
and $e_{3}$ directions ``bend towards'' the $e_{1}$ direction,
we only know the total. In particular, even with positive $e_{1}$
momentum density, a future geodesic in e.g. the $e_{2}$ direction
may ``bend away'' from the $e_{1}$ direction, as long as the future
geodesic in the $e_{3}$ direction ``bends towards'' the $e_{1}$
direction to a greater degree.

\subsection{\label{subsec:energy-conservation-and-divG-vanishing}Local conservation
laws and the divergenceless Einstein tensor}

In this section we show that the vanishing divergence of the SEM tensor,
which physically corresponds to the local conservation of energy and
momentum, has a geometric interpretation in the equivalent vanishing
divergence of the Einstein tensor: namely, it corresponds to the local
conservation of the ``bending'' of space-like geodesics, which ends
up being equivalent to the second Bianchi identity. 

Consider the $0^{\mathrm{th}}$ component of vanishing divergence
\begin{equation}
\begin{aligned}\nabla_{i}T^{0i} & =0\\
\Rightarrow\nabla_{0}T^{00} & =\nabla_{j}T^{0j}.
\end{aligned}
\end{equation}
In continuum mechanics, this corresponds to the physical fact that
the change in particle density in an infinitesimal volume is equal
to the net flux of particles across the faces of the volume's boundary.
In terms of the Einstein tensor, the corresponding relation is
\begin{equation}
\begin{aligned}\nabla_{0}G^{00} & =\nabla_{j}G^{0j}.\end{aligned}
\end{equation}
We arrive at its geometric meaning by substituting in the geometric
meanings of energy density and momentum density: the change in the
total fractional accelerations of infinitesimally separated space-like
geodesics towards each other is equal to the sum of the differences
of fractional accelerations of future space-like geodesics in each
of the three pairs of faces of a cube in the direction orthogonal
to the face. In less precise language, the change in the total bending
of space-like geodesics is equal to the additional amounts that future
geodesics bend towards each other in each space-like direction. 

\begin{figure}[H]
\noindent \begin{centering}
\includegraphics{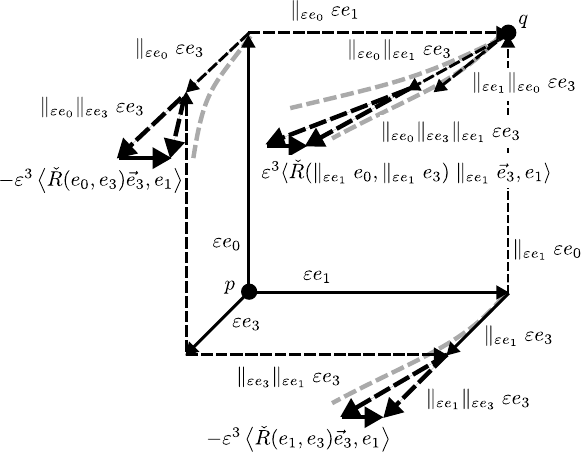}
\par\end{centering}
\caption{\label{fig:second-bianchi}As detailed in the text, the sum of $-\left\langle \parallel_{\varepsilon e_{1}}\check{R}(e_{0},e_{3})\vec{e}_{3},e_{1}\right\rangle $
and $\left\langle \check{R}(\parallel_{\varepsilon e_{1}}e_{0},\parallel_{\varepsilon e_{1}}e_{3})\parallel_{\varepsilon e_{1}}\vec{e}_{3},e_{1}\right\rangle $
is the $e_{1}$ component of incremental convergence between $e_{3}$-directed
geodesics infinitesimally separated in the $e_{1}$ direction due
to being moved forward in time, beyond their initial convergence $\left\langle \check{R}(e_{1},e_{3})\vec{e}_{3},e_{1}\right\rangle $.}
\end{figure}

To see this in more detail, we recall the construction used in the
geometric view of the second Bianchi identity presented in \cite{Marsh-book}
Section 9.2.7, in which we take advantage of the fact that $\check{R}(v,w)\vec{a}$
only depends upon the local value of $\vec{a}$, constructing its
vector field values such that e.g. $\vec{a}\left|_{p+\varepsilon u}\right.=\parallel_{\varepsilon u}(\vec{a}\left|_{p}\right.)$,
so that 
\begin{equation}
\begin{aligned}\varepsilon\nabla_{u}\check{R}(v,w)\vec{a} & =\check{R}(v\left|_{p+\varepsilon u}\right.,w\left|_{p+\varepsilon u}\right.)\vec{a}\left|_{p+\varepsilon u}\right.-\parallel_{\varepsilon u}\check{R}(v,w)\parallel_{\varepsilon u}^{-1}\vec{a}\left|_{p+\varepsilon u}\right.\\
 & =\check{R}(v\left|_{p+\varepsilon u}\right.,w\left|_{p+\varepsilon u}\right.)\parallel_{\varepsilon u}\vec{a}-\parallel_{\varepsilon u}\check{R}(v,w)\vec{a}.
\end{aligned}
\end{equation}
Here, we focus on the second term in (\ref{eq:momentum-density-expression})
\begin{equation}
\nabla_{1}G^{01}=-\nabla_{1}\left\langle \check{R}(e_{0},e_{2})\vec{e}_{2},e_{1}\right\rangle -\nabla_{1}\left\langle \check{R}(e_{0},e_{3})\vec{e}_{3},e_{1}\right\rangle .
\end{equation}
Our cube is already constructed of parallel transports, so that $\vec{e}_{3}\left|_{p+\varepsilon e_{1}}\right.=\parallel_{\varepsilon e_{1}}(\vec{e}_{3}\left|_{p}\right.)$
and $\vec{e}_{0}\left|_{p+\varepsilon e_{1}}\right.=\parallel_{\varepsilon e_{1}}(\vec{e}_{0}\left|_{p}\right.)$,
giving us 
\begin{equation}
\begin{aligned}\varepsilon\nabla_{1}\check{R}(e_{0},e_{3})\vec{e}_{3} & =\check{R}(e_{0}\left|_{p+\varepsilon e_{1}}\right.,e_{3}\left|_{p+\varepsilon e_{1}}\right.)\vec{e}_{3}\left|_{p+\varepsilon e_{1}}\right.\\
 & \phantom{{}=}-\parallel_{\varepsilon e_{1}}\check{R}(e_{0},e_{3})\parallel_{\varepsilon e_{1}}^{-1}\vec{e}_{3}\left|_{p+\varepsilon e_{1}}\right.\\
 & =\check{R}(\parallel_{\varepsilon e_{1}}e_{0},\parallel_{\varepsilon e_{1}}e_{3})\parallel_{\varepsilon e_{1}}\vec{e}_{3}-\parallel_{\varepsilon e_{1}}\check{R}(e_{0},e_{3})\vec{e}_{3}\\
\Rightarrow\varepsilon\left\langle \nabla_{1}\check{R}(e_{0},e_{3})\vec{e}_{3},e_{1}\right\rangle  & =\left\langle \check{R}(\parallel_{\varepsilon e_{1}}e_{0},\parallel_{\varepsilon e_{1}}e_{3})\parallel_{\varepsilon e_{1}}\vec{e}_{3},e_{1}\right\rangle \\
 & \phantom{{}=}-\left\langle \parallel_{\varepsilon e_{1}}\check{R}(e_{0},e_{3})\vec{e}_{3},e_{1}\right\rangle .
\end{aligned}
\end{equation}
The first term, as depicted in Figure \ref{fig:second-bianchi}, can
be viewed as the $e_{1}$ component of the difference between $\parallel_{\varepsilon e_{0}}\parallel_{\varepsilon e_{3}}\parallel_{\varepsilon e_{1}}\varepsilon e_{3}$
and the parallel transport of $\parallel_{\varepsilon e_{1}}\varepsilon e_{3}$
around the far cube face in the opposite direction, which would be
$\parallel_{\varepsilon e_{3}}\parallel_{\varepsilon e_{0}}\parallel_{\varepsilon e_{1}}\varepsilon e_{3}$,
i.e. it is the fractional acceleration of the $e_{3}$-directed geodesic
at $q$ due to being moved forward in time. But this acceleration
is with respect to the geodesic already accelerated by $-\varepsilon^{3}\left\langle \check{R}(e_{1},e_{3})\vec{e}_{3},e_{1}\right\rangle $,
which is the closest thing to a ``straight'' geodesic displaced
in the $e_{1}$ direction. Note that the change in this pre-existing
acceleration due to being moved forward in time is of higher order
in $\varepsilon$ and thus does not change our calculation; neither
does the difference between $\parallel_{\varepsilon e_{0}}\parallel_{\varepsilon e_{1}}\varepsilon e_{3}$
and $\parallel_{\varepsilon e_{1}}\parallel_{\varepsilon e_{0}}\varepsilon e_{3}$.
Thus the first term is an incremental acceleration; to get the total
additional acceleration between adjacent geodesics, we must also add
the second term above, which is the acceleration of the $e_{3}$ directed
geodesic at $p$ due to being moved forward in time (parallel transported
to $q$ to enable addition). Thus the sum $\left\langle \nabla_{1}\check{R}(e_{0},e_{3})\vec{e}_{3},e_{1}\right\rangle $
is the $e_{1}$ component of incremental convergence between $e_{3}$-directed
geodesics infinitesimally separated in the $e_{1}$ direction due
to being moved forward in time, beyond their initial convergence.

As is apparent both from the algebraic form of $\nabla_{1}G^{01}$
and Figure \ref{fig:second-bianchi}, the sum $\nabla_{i}G^{0i}=0$
is equivalent to the second Bianchi identity as depicted in \cite{Marsh-book}
Section 9.2.7, i.e. parallel transporting around each edge of a cube
in opposite directions yields the cancellation of any changes. This
geometric explanation for the vanishing divergence of the Einstein
tensor can similarly be applied to the other components of $\nabla_{i}G^{ji}=0$.

\subsection{\label{subsec:Pressure}Pressure}

\begin{figure}[H]
\noindent \begin{centering}
\includegraphics{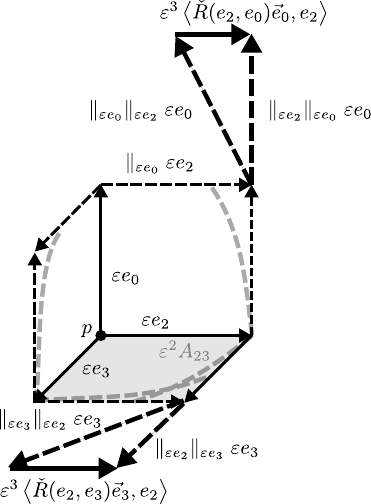}
\par\end{centering}
\caption{\label{fig:pressure}The $e_{1}$ component of normal stress $T^{11}$
is the difference between the convergence of time-like geodesics emanating
from the corners of the cube face $A_{23}$ and the convergence of
space-like geodesics at the outer edges of $A_{23}$.}
\end{figure}

Again using the symmetries of the Riemann curvature tensor, the $e_{1}$
component of normal stress (or pressure if isotropic) is

\begin{equation}
\begin{aligned}T^{11}\propto G^{11} & =-\sum_{\begin{subarray}{c}
m<n\\
m,n\neq1
\end{subarray}}\eta_{mm}\eta_{nn}\left\langle \check{R}(e_{m},e_{n})\vec{e}_{n},e_{m}\right\rangle \\
 & =\left\langle \check{R}(e_{0},e_{2})\vec{e}_{2},e_{0}\right\rangle +\left\langle \check{R}(e_{0},e_{3})\vec{e}_{3},e_{0}\right\rangle -\left\langle \check{R}(e_{2},e_{3})\vec{e}_{3},e_{2}\right\rangle \\
 & =\left\langle \check{R}(e_{2},e_{0})\vec{e}_{0},e_{2}\right\rangle +\left\langle \check{R}(e_{3},e_{0})\vec{e}_{0},e_{3}\right\rangle -\left\langle \check{R}(e_{2},e_{3})\vec{e}_{3},e_{2}\right\rangle .
\end{aligned}
\end{equation}
Again in keeping with the continuum mechanics origin of energy momentum
terminology, the $e_{1}$ normal stress is also called the flux of
$e_{1}$ momentum across the surface $A_{23}$ orthogonal $e_{1}$.
Geometrically, the ``bending'' of future space-like geodesics towards
the past is not a particularly useful concept, so we instead view
the $e_{1}$ normal stress as the fractional acceleration of the infinitesimal
area reduction
\begin{equation}
\begin{aligned}\left\langle \check{R}(e_{2},e_{0})\vec{e}_{0},e_{2}\right\rangle +\left\langle \check{R}(e_{3},e_{0})\vec{e}_{0},e_{3}\right\rangle  & =-\frac{\partial_{0}^{2}A_{23}}{A_{23}}\end{aligned}
\end{equation}
minus the sum of the fractional accelerations of space-like geodesics
at the outer edges of $A_{23}$ towards each other (see Figure \ref{fig:pressure}).

Since the expression for normal stress contains two time-like geodesic
terms and one space-like term, zero pressure in an isotropic configuration
(e.g. the rest frame of dust) means that the acceleration of each
time-like geodesic (gravitational acceleration) is half the acceleration
of each space-like geodesic. In general, we have

\begin{equation}
\begin{aligned}\sum T^{\mathsf{ii}} & \propto2\sum\left\langle \check{R}(e_{\mathsf{j}},e_{0})\vec{e}_{0},e_{\mathsf{j}}\right\rangle -\sum_{\begin{subarray}{c}
\mathsf{i}<\mathsf{j}\end{subarray}}\left\langle \check{R}(e_{\mathsf{i}},e_{\mathsf{j}})\vec{e}_{\mathsf{j}},e_{\mathsf{i}}\right\rangle \\
 & =2\sum\left\langle \check{R}(e_{\mathsf{j}},e_{0})\vec{e}_{0},e_{\mathsf{j}}\right\rangle -G^{00},
\end{aligned}
\end{equation}
and hence the average time-like geodesic acceleration is equal to
the average space-like geodesic acceleration iff $\sum T^{\mathsf{ii}}=T^{00}$,
rather than this sum vanishing as with mass (dust). It is interesting
to note that this equality does hold for null dust, the electromagnetic
field around a massless point charge, or an electromagnetic plane
wave, all of which have total normal stresses equal to their energy
density, and thus equal average (but not isotropic) accelerations
of time-like and space-like geodesics. Combinations of such configurations
can then be constructed to yield accelerations which are both equal
and isotropic. 

\subsection{Stress}

\begin{figure}[H]
\noindent \begin{centering}
\includegraphics{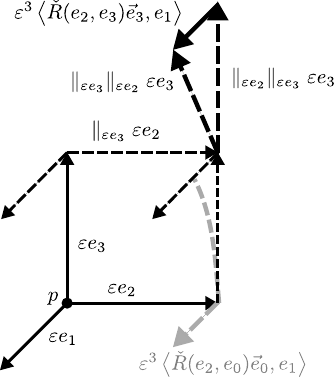}
\par\end{centering}
\caption{\label{fig:stress}The $e_{1}$ component of shear stress in the $e_{2}$
direction $T^{12}$ is the difference between two fractional accelerations
in the $e_{1}$ direction: that of the $e_{2}$-displaced $e_{3}$-geodesic
and that of the $e_{2}$-displaced time-like geodesic.}
\end{figure}

Once again using the symmetries of the Riemann curvature tensor, the
$e_{1}$ component of shear stress in the $e_{2}$ direction is 

\[
\begin{aligned}T^{12}\propto G^{12} & =\sum_{\begin{subarray}{c}
k\neq2\\
k\neq1
\end{subarray}}g_{kk}\left\langle \check{R}(e_{k},e_{2})\vec{e}_{1},e_{k}\right\rangle \\
 & =\sum_{\begin{subarray}{c}
k\neq2\\
k\neq1
\end{subarray}}g_{kk}\left\langle \check{R}(e_{2},e_{k})\vec{e}_{k},e_{1}\right\rangle \\
 & =\left\langle \check{R}(e_{2},e_{3})\vec{e}_{3},e_{1}\right\rangle -\left\langle \check{R}(e_{2},e_{0})\vec{e}_{0},e_{1}\right\rangle .
\end{aligned}
\]
Again in keeping with the continuum mechanics origin of energy momentum
terminology, this is also called the current of $e_{1}$ momentum
in the $e_{2}$ direction, or the flux of $e_{1}$ momentum across
the surface $A_{31}$ orthogonal to $e_{2}$; since $T^{12}=T^{21}$,
$e_{1}$ and $e_{2}$ may be exchanged in these descriptions. Geometrically,
$T^{12}$ is the fractional acceleration of the $e_{2}$-displaced
$e_{3}$-geodesic in the $e_{1}$ direction minus the fractional acceleration
of the $e_{2}$-displaced time-like geodesic in the $e_{1}$ direction
(see Figure \ref{fig:stress}). Zero shear stress thus means that
the $e_{2}$-displaced time-like geodesic ``follows'' the $e_{2}$-displaced
$e_{3}$-geodesic in the $e_{1}$ direction.  

\section{\label{sec:Geometry}Geometric preliminaries}

We next would like to consider the geometric stress, energy, and momentum
detailed in Section \ref{sec:Energy-momentum-as-geodesic-acceleration}
for theories whose actions depend upon the tetrad and spin connection
in addition to the metric. To accomplish this, in this section we
first establish some preliminaries; this is a review of standard material,
but from a particular geometric viewpoint including a level of detail
not always found in other treatments.

\subsection{Lorentz indices and the solder form }

In a choice of coordinates on the spacetime manifold $M$ (or on a
region of $M$), the spacetime connection is 
\begin{equation}
\begin{aligned}\check{\Gamma} & :TM\to gl\left(n,\mathbb{R}\right)\\
\Gamma^{\lambda}{}_{\sigma\mu} & \equiv\mathrm{d}x^{\lambda}\left(\nabla_{\mu}\partial_{\sigma}\right),
\end{aligned}
\end{equation}
where $\partial_{\sigma}$ and $\mathrm{d}x^{\lambda}$ are the coordinate
frame and coframe, also denoted $e_{\sigma}$ and $e^{\lambda}$,
and $\nabla_{\mu}\partial_{\sigma}$ is the difference between the
vector $\partial_{\sigma}$ and its parallel transport in the direction
$\partial_{\mu}$. 

We may also choose an orthonormal frame or tetrad on $M$ (or on a
region of $M$), defining the spin connection 
\begin{equation}
\begin{aligned}\check{\omega} & :TM\to so\left(3,1\right)\\
\omega^{j}{}_{i\mu} & \equiv e^{j}\left(\nabla_{\mu}e_{i}\right),
\end{aligned}
\label{eq:spin-connection}
\end{equation}
where $e_{i}$ and $e^{j}$ are the chosen tetrad and the associated
cotetrad, and $\nabla_{\mu}e_{i}$ is the difference between the vector
$e_{i}$ and its parallel transport in the direction $\partial_{\mu}$;
the component $\omega^{j}{}_{ik}$ then gives this difference in the
direction $e_{k}$.

We use Latin indices to indicate Lorentz indices, i.e. components
in a tetrad or cotetrad; Greek indices are used to indicate coordinate
indices, i.e. components in a coordinate frame or coframe. The ability
to convert between these indices or components is an important tool
when addressing variations of the action. Such a conversion is supplied
by the solder form, or more precisely by its pullback. 

Recall (see e.g. \cite{Marsh-book} Section 10.4.5) that on the orthonormal
frame bundle $FM$, the solder form is a horizontal equivariant $\mathbb{R}^{4}$-valued
1-form $\vec{\theta}_{P}$ which at a point $p=e_{p}\in FM$ projects
its argument down to $M$, and then takes the resulting vector's components
in the basis $e_{p}$. A local trivialization of $FM$ is the selection
of an identity section as a local tetrad, and the pullback of $\vec{\theta}_{P}$
by this identity section can then be viewed as a frame-dependent $\mathbb{R}^{4}$-valued
1-form on $M$ which returns the components of its argument in this
local tetrad, i.e. 
\begin{equation}
\begin{aligned}\theta:TM & \to\mathbb{R}^{4}\\
v & \mapsto e^{i}\left(v\right).
\end{aligned}
\end{equation}
In the present context, this 1-form is written in a local coordinate
frame on $M$ as a mixed index tensor (which we note is an extension
of the usual definition of a tensor)
\begin{equation}
\begin{aligned}e_{\mu}^{i}:TM & \to\mathbb{R}^{4}\\
v^{\mu} & \mapsto e_{\mu}^{i}v^{\mu}.
\end{aligned}
\end{equation}
This mixed index tensor can also be viewed as a multilinear mapping
$e$ whose component array values are defined by applying it to basis
vectors in different frames:
\begin{equation}
\begin{aligned}e\left(\varphi,v\right) & \equiv\varphi\left(v\right),\\
e_{\mu}^{i} & \equiv e\left(e^{i},e_{\mu}\right)\\
 & =e^{i}\left(e_{\mu}\right).
\end{aligned}
\label{eq:e-as-multilinear-mapping}
\end{equation}

Now, the value of $e_{\mu}^{i}$ at a given point may be viewed as
either the components of $e_{\mu}=\partial_{\mu}$ in the frame $e_{i}$,
or the components of $e^{i}$ in the coframe $e^{\mu}=\mathrm{d}x^{\mu}$,
i.e.
\begin{equation}
\begin{aligned}e_{\mu}^{i} & =e^{i}(e_{\mu})=\left(e_{\mu}\right)^{i}=\left(e^{i}\right)_{\mu}.\end{aligned}
\end{equation}
We may therefore use it to convert between coordinate and Lorentz
indices, and to form other tensors with mixed indices:
\begin{equation}
\begin{aligned}v^{i} & =e^{i}\left(v^{\mu}e_{\mu}\right)=e_{\mu}^{i}v^{\mu},\\
\varphi_{\mu} & =e_{\mu}\left(\varphi_{i}e^{i}\right)=e_{\mu}^{i}\varphi_{i},\\
\Rightarrow e_{\mu}^{i}A^{\mu\nu} & \equiv e_{\mu}^{i}\left(v^{\mu}\otimes w^{\nu}\right)=v^{i}\otimes w^{\nu}=A^{i\nu}.
\end{aligned}
\end{equation}
The indices of such tensors may be lowered with $g_{\mu\nu}$ and
$\eta_{ij}$ and raised with $g^{\mu\nu}$ and $\eta^{ij}$, including
those of $e_{\mu}^{i}$ itself; this allows us to obtain $e_{i}^{\mu}$,
which may be viewed as the components of $e_{i}$ in the frame $e_{\mu}$
or the components of $e^{\mu}$ in the coframe $e^{i}$. It also allows
us to e.g. use $e_{i\mu}$ to convert the coordinate frame components
of a tangent vector on $M$ to the cotetrad components of the dual
covector.

As we did with $e$, we may view the mixed index tensor $A$ as a
multilinear mapping whose component array values are defined by applying
it to basis vectors in different frames; this allows us to take the
covariant derivative of mixed index tensors by applying the two connections,
e.g. 
\begin{equation}
\begin{aligned}\nabla_{\lambda}A^{i\nu} & =\partial_{\lambda}A^{i\nu}+\Gamma^{\nu}{}_{\mu\lambda}A^{i\mu}+\omega^{i}{}_{j\lambda}A^{j\nu}.\end{aligned}
\end{equation}
The covariant derivative can then be applied to $e_{i}^{\mu}$ itself,
yielding
\begin{equation}
\begin{aligned}\nabla_{\lambda}e_{i}^{\mu} & =\partial_{\lambda}e_{i}^{\mu}+\Gamma^{\mu}{}_{\nu\lambda}e_{i}^{\nu}-\omega^{j}{}_{i\lambda}e_{j}^{\mu}\\
 & =\partial_{\lambda}e_{i}^{\mu}+\Gamma^{\mu}{}_{\nu\lambda}e_{i}^{\nu}-\left(\nabla_{\lambda}e_{i}\right)^{\mu}\\
 & =\partial_{\lambda}e_{i}^{\mu}+\Gamma^{\mu}{}_{\nu\lambda}e_{i}^{\nu}-\partial_{\lambda}e_{i}^{\mu}-\Gamma^{\mu}{}_{\nu\lambda}e_{i}^{\nu}\\
 & =0,
\end{aligned}
\end{equation}
where we have used (\ref{eq:spin-connection}) and recall (see \cite{Marsh-book}
Section 9.1.5) that when using index notation, the covariant derivative
of a vector is

\begin{equation}
\begin{aligned}\nabla_{\lambda}v^{\mu} & \equiv\left(\nabla_{\lambda}v\right)^{\mu}\\
 & =\partial_{\lambda}v^{\mu}-\Gamma^{\mu}{}_{\nu\lambda}v^{\nu}.
\end{aligned}
\label{eq:del-v-expressions}
\end{equation}
Note that using the Leibniz rule for the covariant derivative of a
vector multiplied by a function, we have 

\begin{equation}
\begin{aligned}\nabla_{\lambda}v^{\mu} & =\left(\nabla_{\lambda}v\right)^{\mu}\\
 & =\left(\nabla_{\lambda}\left(e_{i}v^{i}\right)\right)^{\mu}\\
 & =\left(e_{i}\partial_{\lambda}v^{i}\right)^{\mu}+\left(v^{i}\nabla_{\lambda}e_{i}\right)^{\mu}\\
 & =\left(e_{i}\right)^{\mu}\partial_{\lambda}v^{i}+v^{i}\left(\nabla_{\lambda}e_{i}\right)^{\mu}\\
 & =e_{i}^{\mu}\partial_{\lambda}v^{i}+v^{i}\omega^{\mu}{}_{i\lambda}\\
 & =e_{i}^{\mu}\left(\partial_{\lambda}v^{i}+\omega^{i}{}_{j\lambda}v^{j}\right)\\
 & =e_{i}^{\mu}\nabla_{\lambda}v^{i},
\end{aligned}
\end{equation}

\noindent where we have used (\ref{eq:spin-connection}) and relabeled
dummy indices. This confirms that the Leibniz rule continues to apply
to mixed index tensors, since the above may be written as the covariant
derivative of such a tensor contracted with a vector: 

\begin{equation}
\begin{aligned}\nabla_{\lambda}v^{\mu} & =\left(\nabla_{\lambda}v\right)^{\mu}\\
 & =\left(\nabla_{\lambda}\left(e_{i}v^{i}\right)\right)^{\mu}\\
 & =\nabla_{\lambda}\left(e_{i}^{\mu}v^{i}\right)\\
 & =v^{i}\nabla_{\lambda}e_{i}^{\mu}+e_{i}^{\mu}\nabla_{\lambda}v^{i}\\
 & =e_{i}^{\mu}\nabla_{\lambda}v^{i}.
\end{aligned}
\end{equation}

\subsection{Some expressions using Lorentz indices}

We define

\begin{equation}
\begin{aligned}\partial & \in\left\{ \partial_{\lambda},\partial_{k},\delta\right\} \end{aligned}
\end{equation}
to be any derivative of functions on $M$, whether due to an infinitesimal
change in a coordinate frame direction $\partial_{\lambda}$, or a
tetrad direction $\partial_{k}$, or in the frame itself due to a
variation $\delta e_{i}^{\mu}$, which is not to be confused with
the Kronecker delta $\delta_{\nu}^{\mu}$. Since the components of
$e$ may be viewed as functions, we then have
\begin{equation}
\begin{aligned}e_{i}^{\mu}e_{\mu}^{j} & =e_{i}^{\mu}g_{\mu\nu}\eta^{kj}e_{k}^{\nu}=\eta^{kj}\left\langle e_{i},e_{k}\right\rangle =\delta_{i}^{j}\\
\Rightarrow\partial\left(e_{i}^{\mu}e_{\mu}^{j}\right) & =0\\
\Rightarrow e_{\mu}^{j}\partial e_{i}^{\mu} & =-e_{i}^{\mu}\partial e_{\mu}^{j}\\
\Rightarrow e_{\nu}^{i}e_{\mu}^{j}\partial e_{i}^{\mu} & =-e_{\nu}^{i}e_{i}^{\mu}\partial e_{\mu}^{j}\\
 & =-\delta_{\nu}^{\mu}\partial e_{\mu}^{j}\\
\Rightarrow\partial e_{\nu}^{j} & =-e_{\nu}^{i}e_{\mu}^{j}\partial e_{i}^{\mu},
\end{aligned}
\end{equation}
providing a way to swap Lorentz and coordinate indices in the tetrad
derivative. Similarly, we have
\begin{equation}
\begin{aligned}e_{j}^{\nu}e_{\mu}^{j}\partial e_{i}^{\mu} & =-e_{j}^{\nu}e_{i}^{\mu}\partial e_{\mu}^{j}\\
\Rightarrow\partial e_{i}^{\nu} & =-e_{j}^{\nu}e_{i}^{\mu}\partial e_{\mu}^{j}.
\end{aligned}
\end{equation}

We also can see that $g$ and $\eta$ may be written
\begin{align}
\begin{aligned}g_{\mu\nu} & =e_{\mu}^{i}e_{\nu}^{j}\eta_{ij}\\
 & =e_{\mu}^{i}e_{i\nu},\\
\eta_{ij} & =e_{i}^{\mu}e_{j}^{\nu}g_{\mu\nu}\\
 & =e_{i}^{\mu}e_{\mu j}\\
\Rightarrow\partial\left(g_{\mu\nu}\right) & =\eta_{ij}\left(\partial e_{\mu}^{i}e_{\nu}^{j}+e_{\mu}^{i}\partial e_{\nu}^{j}\right),\\
\partial\left(e_{i}^{\mu}e_{\mu j}\right) & =0\\
\Rightarrow e_{i}^{\mu}\partial e_{\mu j} & =-e_{\mu j}\partial e_{i}^{\mu}.
\end{aligned}
\end{align}
Recalling that 

\begin{equation}
\begin{aligned}\delta\left(\sqrt{g}\right) & =\frac{1}{2}\sqrt{g}g^{\mu\nu}\delta g_{\mu\nu}=-\frac{1}{2}\sqrt{g}g_{\mu\nu}\delta g^{\mu\nu}.\end{aligned}
\label{eq:variation-of-sqrt-g}
\end{equation}
and using the above expression for the metric it is also not hard
to arrive at 
\begin{align}
\begin{aligned}\delta\left(\sqrt{g}\right) & =\sqrt{g}e_{i}^{\mu}\delta e_{\mu}^{i}\\
 & =-\sqrt{g}e_{\mu}^{i}\delta e_{i}^{\mu}.
\end{aligned}
\end{align}

Recalling the expression for torsion in an arbitrary frame and the
relation $e_{\mu}^{j}\partial_{\nu}e_{i}^{\mu}=-e_{i}^{\mu}\partial_{\nu}e_{\mu}^{j}$
above, we may write an expression for the torsion tensor in terms
of $e$ and $\omega$ as
\begin{equation}
\begin{aligned}T^{i}{}_{jk} & =\omega^{i}{}_{kj}-\omega^{i}{}_{jk}-[e_{j},e_{k}]^{i}\\
 & =\omega^{i}{}_{kj}-\omega^{i}{}_{jk}-e_{\sigma}^{i}e_{j}^{\rho}\partial_{\rho}e_{k}^{\sigma}+e_{\sigma}^{i}e_{k}^{\rho}\partial_{\rho}e_{j}^{\sigma}\\
 & =\omega^{i}{}_{kj}-\omega^{i}{}_{jk}+e_{k}^{\sigma}e_{j}^{\rho}\partial_{\rho}e_{\sigma}^{i}-e_{j}^{\sigma}e_{k}^{\rho}\partial_{\rho}e_{\sigma}^{i}\\
\Rightarrow T^{\lambda}{}_{\mu\nu} & =e_{i}^{\lambda}e_{\mu}^{j}e_{\nu}^{k}T^{i}{}_{jk}\\
 & =e_{i}^{\lambda}e_{\nu}^{k}\omega^{i}{}_{k\mu}-e_{i}^{\lambda}e_{\mu}^{j}\omega^{i}{}_{j\nu}+e_{i}^{\lambda}\delta_{\mu}^{\rho}\delta_{\nu}^{\sigma}\partial_{\rho}e_{\sigma}^{i}-e_{i}^{\lambda}\delta_{\mu}^{\sigma}\delta_{\nu}^{\rho}\partial_{\rho}e_{\sigma}^{i}\\
 & =e_{i}^{\lambda}\left(e_{\nu}^{j}\omega^{i}{}_{j\mu}-e_{\mu}^{j}\omega^{i}{}_{j\nu}+\partial_{\mu}e_{\nu}^{i}-\partial_{\nu}e_{\mu}^{i}\right).
\end{aligned}
\label{eq:torsion-expansion}
\end{equation}
Note the sign reversal of the frame derivative terms on the third
line due to the conversion of the indices of the frame component under
the derivative. 

\subsection{Varying parallel transport}

We view parallel transport as the intrinsic quantity to be varied
on $M$. When addressing actions which are expressed in terms of the
metric (via inner products and the volume element) and the spacetime
connection (via the covariant derivative and the scalar curvature),
we may accomplish this by varying the connection $\Gamma$ in a chosen
frame. A general $\Gamma$ whose parallel transport has holonomy group
$SO(3,1)$ is associated with a unique Lorentzian metric (up to a
choice of units) $g$, and may be written as 
\begin{align}
\begin{aligned}\Gamma^{\lambda}{}_{\mu\nu} & =\overline{\Gamma}^{\lambda}{}_{\mu\nu}+K^{\lambda}{}_{\mu\nu},\\
\omega^{i}{}_{jk} & =\overline{\omega}^{i}{}_{jk}+K^{i}{}_{jk},
\end{aligned}
\end{align}
where $\overline{\Gamma}$ and $\overline{\omega}$ are the torsionless
connection and spin connection associated with $g$, and $K_{\lambda\mu\nu}$
is the contorsion tensor
\begin{align}
\begin{aligned}K_{\lambda\mu\nu} & =\frac{1}{2}\left(T_{\mu\lambda\nu}+T_{\nu\lambda\mu}-T_{\lambda\mu\nu}\right),\end{aligned}
\label{eq:K-in-terms-of-T}
\end{align}
which is anti-symmetric in its first two indices. We may therefore
view the variation of $\Gamma$ as independent variations of $\overline{\Gamma}$
and $K$. The variation of $\overline{\Gamma}$ while holding $K$
constant will induce a variation $\delta g$, and this may be reversed
to view $\delta\overline{\Gamma}$ as being induced by a variation
of $g$, making $g$ and $K$ the independent variations. The variation
of $K$ while holding $\overline{\Gamma}$ and therefore $g$ constant
means that we may raise and lower indices at will.

It is important to note that independently varying $g$ and $K$ is
not equivalent to independently varying $g$ and $T$. It is clear
from the definition of $K$ in terms of $T$ that it is dependent
upon $g$ via the lowered indices, so that varying $g$ while holding
$K$ constant results in altered torsion, and varying $g$ while holding
$T$ constant by necessity alters $K$. However, varying $K$ while
holding $g$ constant is equivalent to varying $T$ while holding
$g$ constant; perhaps for this reason, these two different variations
are sometimes described interchangeably.

Now, when addressing actions such as that of Dirac theory, which are
expressed in terms of the tetrad (via the Dirac matrices) and the
spin connection (via the covariant derivative and scalar curvature),
we must instead vary the parallel transport by using the Sciama-Kibble
formalism, which varies the tetrad $e$ and the spin connection $\omega$.
The tetrad uniquely determines a metric, and in terms of this tetrad,
the spin connection $\omega$ is a matrix of 1-forms $\omega^{i}{}_{k}$,
which may be kept constant while varying $e$; thus we may vary the
parallel transport via independent variations of $e$ and $\omega^{i}{}_{k}$. 

As with the variation of $g$ and $K$, it is important to note that
independently varying $e$ and $\omega$ is not equivalent to independently
varying $g$ and $T$. For example, if we limit our variation of $e$
to a ``rotation'' at each point (a Lorentz transformation), then
holding $\omega^{i}{}_{k}$ constant results in altered torsion, while
such a variation of $e$ while holding torsion constant by necessity
alters $\omega^{i}{}_{k}$. However, varying $\omega^{i}{}_{k}$ while
holding $e$ constant is equivalent to varying $T$ while holding
$g$ constant, just as varying $K$ while holding $g$ constant is.

Using the first expression in (\ref{eq:torsion-expansion}), we may
write $K$ in terms of $e$ and $\omega$; it is not hard to see that
due to the anti-symmetry of the first two indices of $\omega_{ijk}$,
there are two pairs of cancellations resulting in
\begin{equation}
\begin{aligned}K^{i}{}_{j\mu} & =\omega^{i}{}_{j\mu}+f\left(e\right)\\
\Rightarrow\frac{\partial K^{i}{}_{j\mu}}{\partial\omega^{i}{}_{j\mu}} & =1,
\end{aligned}
\label{eq:var-K-same-as-var-omega}
\end{equation}
where $f$ is some function of the tetrad which leaves the derivative
unaffected. From a variational standpoint this makes sense: the tetrad
is associated with a unique metric, which in turn is associated with
a unique torsionless Levi-Civita spin connection; therefore any variation
of the spin connection while keeping the tetrad constant directly
varies the torsion, and hence the contorsion.

\section{\label{sec:SEM-tensors}Stress energy momentum tensors}

In this section we define various SEM tensors and obtain relationships
between them, for a generic action defined as
\begin{equation}
\begin{aligned}S & \equiv S_{\mathrm{M}}+S_{\mathrm{G}}\\
 & \equiv\int\mathfrak{L}_{\mathrm{M}}\mathrm{d}^{4}x+\frac{1}{2\kappa}\int R\sqrt{g}\mathrm{d}^{4}x,
\end{aligned}
\label{eq:total-action}
\end{equation}
where $S_{\mathrm{M}}$ is the matter action, $S_{\mathrm{G}}$ is
the gravitational action, $\mathfrak{L}_{\mathrm{M}}$ is the matter
Lagrangian density, $R$ is the scalar curvature, and 
\begin{equation}
\sqrt{g}\equiv\sqrt{\left|\det\left(g_{\mu\nu}\right)\right|}.
\end{equation}
We will denote the variational derivative of the action with respect
to e.g. the metric by $\delta S(g)$, while matter and gravitational
actions under the assumption of zero torsion will be denoted with
an overbar, e.g. $\mathfrak{L}_{\overbar{\mathrm{M}}}$.

These are classic results, presented here from a geometric viewpoint
in detail and in the mostly pluses signature.

\subsection{The Hilbert SEM tensor}

The definition of the \textbf{Hilbert SEM tensor} $\overline{T}$
and its density $\overline{\mathfrak{T}}$ is 
\begin{equation}
\begin{aligned}\delta S_{\mathrm{M}}\left(g\right) & =\frac{1}{2}\int\left(\frac{\partial\mathfrak{L}_{\mathrm{M}}}{\partial g_{\mu\nu}}+\frac{\partial\mathfrak{L}_{\mathrm{M}}}{\partial g_{\nu\mu}}\right)\delta g_{\mu\nu}\mathrm{d}^{4}x\\
 & \equiv\frac{1}{2}\int\overline{\mathfrak{T}}^{\mu\nu}\delta g_{\mu\nu}\mathrm{d}^{4}x\\
 & \equiv\frac{1}{2}\int\overline{T}^{\mu\nu}\sqrt{g}\delta g_{\mu\nu}\mathrm{d}^{4}x\\
 & =-\frac{1}{2}\int\overline{T}_{\mu\nu}\delta g^{\mu\nu}\mathrm{d}V,
\end{aligned}
\label{eq:hilbert-SEM-definition}
\end{equation}
where $\mathrm{d}V=\sqrt{g}\mathrm{d}^{4}x$, we explicitly vary the
metric components symmetrically in order to reflect the variation
of the underlying geometric quantity, ensuring $\overline{T}$ and
$\overline{\mathfrak{T}}$ are symmetric, and we recall that 
\begin{equation}
\delta g^{\lambda\sigma}=-g^{\lambda\mu}g^{\nu\sigma}\delta g_{\mu\nu}.
\end{equation}

If we assume $K=0$ as in standard general relativity, the variation
of the gravitational action is
\begin{equation}
\begin{aligned}2\kappa\delta S_{\overbar{\mathrm{G}}}\left(g\right) & =\int\delta\left(\overline{R}\sqrt{g}\right)\mathrm{d}^{4}x\\
 & =\int\left(\sqrt{g}\delta\overline{R}+\overline{R}\delta\left(\sqrt{g}\right)\right)\mathrm{d}^{4}x\\
 & =\int\left(\overline{R}_{\mu\nu}\delta g^{\mu\nu}+g^{\mu\nu}\delta\overline{R}_{\mu\nu}-\overline{R}\frac{1}{2}g_{\mu\nu}\delta g^{\mu\nu}\right)\mathrm{d}V\\
 & =\int\overline{G}_{\mu\nu}\delta g^{\mu\nu}\mathrm{d}V+\int g^{\mu\nu}\delta\overline{R}_{\mu\nu}\mathrm{d}V,
\end{aligned}
\label{eq:grav-variation-g}
\end{equation}
where $\overline{R}$ and $\overline{R}_{\mu\nu}$ are the torsionless
scalar curvature and Ricci tensor, and we arrive at the torsionless
Einstein tensor 
\begin{equation}
\overline{G}_{\mu\nu}\equiv\overline{R}_{\mu\nu}-\frac{1}{2}\overline{R}g_{\mu\nu}
\end{equation}
by recalling (\ref{eq:variation-of-sqrt-g}).

Regarding the second integral, we use the expression for the curvature
in terms of the connection in a coordinate frame (see e.g. \cite{Marsh-book}
Section 9.2.5) to write
\begin{equation}
\begin{aligned}\delta\overline{R}_{\mu\nu} & =\delta\overline{R}^{\lambda}{}_{\mu\lambda\nu}\\
 & =\delta\left(\partial_{\lambda}\overline{\Gamma}^{\lambda}{}_{\mu\nu}-\partial_{\nu}\overline{\Gamma}^{\lambda}{}_{\mu\lambda}+\overline{\Gamma}^{\lambda}{}_{\sigma\lambda}\overline{\Gamma}^{\sigma}{}_{\mu\nu}-\overline{\Gamma}^{\lambda}{}_{\sigma\nu}\overline{\Gamma}^{\sigma}{}_{\mu\lambda}\right)\\
 & =\partial_{\lambda}\delta\overline{\Gamma}^{\lambda}{}_{\mu\nu}-\partial_{\nu}\delta\overline{\Gamma}^{\lambda}{}_{\mu\lambda}+\delta\overline{\Gamma}^{\lambda}{}_{\sigma\lambda}\overline{\Gamma}^{\sigma}{}_{\mu\nu}+\overline{\Gamma}^{\lambda}{}_{\sigma\lambda}\delta\overline{\Gamma}^{\sigma}{}_{\mu\nu}\\
 & \phantom{{}=}-\delta\overline{\Gamma}^{\lambda}{}_{\sigma\nu}\overline{\Gamma}^{\sigma}{}_{\mu\lambda}-\overline{\Gamma}^{\lambda}{}_{\sigma\nu}\delta\overline{\Gamma}^{\sigma}{}_{\mu\lambda}\\
 & =\partial_{\lambda}\delta\overline{\Gamma}^{\lambda}{}_{\mu\nu}+\overline{\Gamma}^{\lambda}{}_{\sigma\lambda}\delta\overline{\Gamma}^{\sigma}{}_{\mu\nu}-\overline{\Gamma}^{\sigma}{}_{\mu\lambda}\delta\overline{\Gamma}^{\lambda}{}_{\sigma\nu}-\overline{\Gamma}^{\lambda}{}_{\sigma\nu}\delta\overline{\Gamma}^{\sigma}{}_{\mu\lambda}\\
 & \phantom{{}=}-\left(\partial_{\nu}\delta\overline{\Gamma}^{\lambda}{}_{\mu\lambda}-\overline{\Gamma}^{\sigma}{}_{\mu\nu}\delta\overline{\Gamma}^{\lambda}{}_{\sigma\lambda}\right).
\end{aligned}
\end{equation}
Recall that while $\overline{\Gamma}$ is not a tensor, the difference
$\delta\overline{\Gamma}$ is; for the last term on the penultimate
line, we may therefore relabel dummy indices and use the symmetric
lower indices of the torsionless connection to arrive at the Palatini
identity 
\begin{equation}
\begin{aligned}\delta\overline{R}_{\mu\nu} & =\overline{\nabla}_{\lambda}\left(\delta\overline{\Gamma}^{\lambda}{}_{\mu\nu}\right)-\overline{\nabla}_{\nu}\left(\delta\overline{\Gamma}^{\lambda}{}_{\mu\lambda}\right)\\
\Rightarrow g^{\mu\nu}\delta\overline{R}_{\mu\nu} & =g^{\mu\nu}\overline{\nabla}_{\lambda}\left(\delta\overline{\Gamma}^{\lambda}{}_{\mu\nu}\right)-g^{\mu\nu}\overline{\nabla}_{\nu}\left(\delta\overline{\Gamma}^{\lambda}{}_{\mu\lambda}\right)\\
 & =\overline{\nabla}_{\lambda}\left(g^{\mu\nu}\delta\overline{\Gamma}^{\lambda}{}_{\mu\nu}\right)-\overline{\nabla}_{\nu}\left(g^{\mu\nu}\delta\overline{\Gamma}^{\lambda}{}_{\mu\lambda}\right)\\
 & =\overline{\nabla}_{\lambda}\left(g^{\mu\nu}\delta\overline{\Gamma}^{\lambda}{}_{\mu\nu}-g^{\mu\lambda}\delta\overline{\Gamma}^{\nu}{}_{\mu\nu}\right).
\end{aligned}
\end{equation}
The second term in the last line of (\ref{eq:grav-variation-g}) thus
vanishes due to the divergence theorem, since we assume that $\overline{\Gamma}$
(the underlying quantity being varied) vanishes at the boundary, and
our EOM for $K=0$ are the Einstein field equations
\begin{equation}
\overline{G}^{\mu\nu}=\kappa\overline{T}^{\mu\nu}.
\end{equation}
Under these EOM, since $\overline{G}$ is divergenceless, so is $\overline{T}$.

We now show that if we include torsion in the scalar curvature of
the gravitational action, then for any matter Lagrangian $\mathfrak{L}_{\overbar{\mathrm{M}}}$
which does not depend upon $K$, the variation of $K$ results in
the EOM $K=0$, and therefore the Einstein field equations remain
valid. We first express the scalar curvature in terms of $\overline{\Gamma}$,
$g$, and $K$ by writing
\begin{equation}
\begin{aligned}R_{\mu\nu} & =\partial_{\lambda}\Gamma^{\lambda}{}_{\mu\nu}-\partial_{\nu}\Gamma^{\lambda}{}_{\mu\lambda}+\Gamma^{\lambda}{}_{\sigma\lambda}\Gamma^{\sigma}{}_{\mu\nu}-\Gamma^{\lambda}{}_{\sigma\nu}\Gamma^{\sigma}{}_{\mu\lambda}\\
 & =\overline{R}_{\mu\nu}+\partial_{\lambda}K^{\lambda}{}_{\mu\nu}-\partial_{\nu}K^{\lambda}{}_{\mu\lambda}+K^{\lambda}{}_{\sigma\lambda}K^{\sigma}{}_{\mu\nu}-K^{\lambda}{}_{\sigma\nu}K^{\sigma}{}_{\mu\lambda}\\
 & \phantom{{}=}+\overline{\Gamma}^{\lambda}{}_{\sigma\lambda}K^{\sigma}{}_{\mu\nu}-\overline{\Gamma}^{\lambda}{}_{\sigma\nu}K^{\sigma}{}_{\mu\lambda}+K^{\lambda}{}_{\sigma\lambda}\overline{\Gamma}^{\sigma}{}_{\mu\nu}-K^{\lambda}{}_{\sigma\nu}\overline{\Gamma}^{\sigma}{}_{\mu\lambda}\\
 & =\overline{R}_{\mu\nu}+\partial_{\lambda}K^{\lambda}{}_{\mu\nu}+\overline{\Gamma}^{\lambda}{}_{\sigma\lambda}K^{\sigma}{}_{\mu\nu}-\overline{\Gamma}^{\sigma}{}_{\nu\lambda}K^{\lambda}{}_{\mu\sigma}-\overline{\Gamma}^{\sigma}{}_{\mu\lambda}K^{\lambda}{}_{\sigma\nu}\\
 & \phantom{{}=}-\partial_{\nu}K^{\lambda}{}_{\mu\lambda}+\overline{\Gamma}^{\sigma}{}_{\mu\nu}K^{\lambda}{}_{\sigma\lambda}+K^{\lambda}{}_{\sigma\lambda}K^{\sigma}{}_{\mu\nu}-K^{\lambda}{}_{\sigma\nu}K^{\sigma}{}_{\mu\lambda}\\
 & =\overline{R}_{\mu\nu}+\overline{\nabla}_{\lambda}K^{\lambda}{}_{\mu\nu}-\overline{\nabla}_{\nu}K^{\lambda}{}_{\mu\lambda}+K^{\lambda}{}_{\sigma\lambda}K^{\sigma}{}_{\mu\nu}-K^{\lambda}{}_{\sigma\nu}K^{\sigma}{}_{\mu\lambda}\\
\Rightarrow R & =g^{\mu\nu}R_{\mu\nu}\\
 & =\overline{R}+\overline{\nabla}_{\lambda}K^{\lambda\nu}{}_{\nu}-\overline{\nabla}_{\nu}K^{\lambda\nu}{}_{\lambda}+K^{\lambda}{}_{\sigma\lambda}K^{\sigma\nu}{}_{\nu}-K^{\lambda}{}_{\sigma\nu}K^{\sigma\nu}{}_{\lambda}\\
 & =\overline{R}+2\overline{\nabla}_{\lambda}\left(g^{\mu\nu}K^{\lambda}{}_{\mu\nu}\right)-g^{\mu\nu}\left(K^{\lambda}{}_{\mu\lambda}K^{\sigma}{}_{\nu\sigma}-K^{\lambda}{}_{\mu\sigma}K^{\sigma}{}_{\nu\lambda}\right),
\end{aligned}
\label{eq:expansion-of-R}
\end{equation}
where in the fourth term of the fourth line we have relabeled dummy
indices and use the symmetry of the lower indices of the torsionless
connection to find that $\overline{\Gamma}^{\lambda}{}_{\sigma\nu}K^{\sigma}{}_{\mu\lambda}=\overline{\Gamma}^{\sigma}{}_{\nu\lambda}K^{\lambda}{}_{\mu\sigma}$,
while in the last line we have relabeled dummy indices and used the
anti-symmetry of the first two indices in $K$. Having isolated the
terms in $K$, we may now vary it to find that
\begin{equation}
\begin{aligned}2\kappa\delta S_{\mathrm{G}}\left(K_{\sigma\lambda\nu}\right) & =\int2\overline{\nabla}_{\lambda}\left(\delta K^{\lambda\nu}{}_{\nu}\right)\mathrm{d}V+\int\left(\delta\left(K^{\lambda}{}_{\sigma\lambda}K^{\sigma\nu}{}_{\nu}\right)-\delta\left(K^{\lambda}{}_{\sigma\nu}\delta K^{\sigma\nu}{}_{\lambda}\right)\right)\mathrm{d}V\\
 & =\int\left(K^{\lambda}{}_{\sigma\lambda}\delta K^{\sigma\nu}{}_{\nu}+K^{\sigma\nu}{}_{\nu}\delta K^{\lambda}{}_{\sigma\lambda}-K^{\lambda}{}_{\sigma\nu}\delta K^{\sigma\nu}{}_{\lambda}-K^{\sigma\nu}{}_{\lambda}\delta K^{\lambda}{}_{\sigma\nu}\right)\mathrm{d}V\\
 & =\int\left(K^{\lambda}{}_{\sigma\lambda}\delta K^{\sigma\nu}{}_{\nu}+K^{\nu}{}_{\sigma\nu}\delta K^{\sigma\lambda}{}_{\lambda}-K^{\nu}{}_{\sigma\lambda}\delta K^{\sigma\lambda}{}_{\nu}+K^{\sigma\nu\lambda}\delta K_{\sigma\lambda\nu}\right)\mathrm{d}V\\
 & =2\int\left(K^{\mu\sigma}{}_{\mu}g^{\lambda\nu}+K^{\sigma\nu\lambda}\right)\delta K_{\sigma\lambda\nu}\mathrm{d}V\\
\Rightarrow2\kappa\delta S_{\mathrm{G}}\left(K\right) & =\kappa\delta S\left(K_{\sigma\lambda\nu}\right)-\kappa\delta S\left(K_{\lambda\sigma\nu}\right)\\
 & =\int\left(K^{\mu\sigma}{}_{\mu}g^{\lambda\nu}+K^{\sigma\nu\lambda}-K^{\mu\lambda}{}_{\mu}g^{\sigma\nu}-K^{\lambda\nu\sigma}\right)\delta K_{\sigma\lambda\nu}\mathrm{d}V,
\end{aligned}
\label{eq:vary-K-in-grav-action}
\end{equation}
where the first integral vanishes by the divergence theorem since
we assume the variation vanishes on the integration boundary, we have
again relabeled dummy indices and used the anti-symmetry of the first
two indices in $K$, and in the last line we have anti-symmetrized
to reflect variation of the underlying geometric quantity. Assuming
the matter Lagrangian does not depend on $K$, the integrand must
vanish; taking the trace, we see that
\begin{equation}
\begin{aligned}0 & =g_{\sigma\nu}\left(K^{\mu\sigma}{}_{\mu}g^{\lambda\nu}-K^{\mu\lambda}{}_{\mu}g^{\sigma\nu}+K^{\sigma\nu\lambda}-K^{\lambda\nu\sigma}\right)\\
 & =K^{\mu\lambda}{}_{\mu}-4K^{\mu\lambda}{}_{\mu}+K_{\nu}{}^{\nu\lambda}-K^{\lambda\nu}{}_{\nu}\\
 & =-2K^{\mu\lambda}{}_{\mu}\\
\Rightarrow K^{\mu\lambda}{}_{\mu} & =0,
\end{aligned}
\end{equation}
where again we have relabeled dummy indices and used the anti-symmetry
of the first two indices in $K$, by which the third term on the second
line vanishes. But again using the anti-symmetric indices, the remaining
terms are 
\begin{align}
\int\left(K^{\sigma\nu\lambda}-K^{\lambda\nu\sigma}\right)\delta K_{\sigma\lambda\nu}\mathrm{d}V & =\int\left(K^{\nu\lambda\sigma}-K^{\nu\sigma\lambda}\right)\delta K_{\sigma\lambda\nu}\mathrm{d}V.
\end{align}
For this integrand to vanish, $K$ must be symmetric in the last two
indices; but since it is also anti-symmetric in the first two indices,
the entire tensor must vanish. Thus $K=0$ for any matter Lagrangian
which does not depend on $K$, and the Einstein field equations remain
valid, justifying our use of the scalar curvature including torsion
as the gravitational action in (\ref{eq:total-action}).

\subsection{The spin SEM tensor}

We now consider the more general case in which the matter Lagrangian
does depend upon the contorsion, arriving at Einstein-Cartan theory.
We first address the \textbf{spin tensor} and its density; they are
defined analogously (but with opposite sign) to the Hilbert SEM tensor
in (\ref{eq:hilbert-SEM-definition}) by
\begin{equation}
\begin{aligned}\delta S_{\mathrm{M}}\left(K\right) & =\frac{1}{2}\int\left(\frac{\partial\mathfrak{L}_{\mathrm{M}}}{\partial K_{\sigma\lambda\nu}}-\frac{\partial\mathfrak{L}_{\mathrm{M}}}{\partial K_{\lambda\sigma\nu}}\right)\delta K_{\sigma\lambda\nu}\mathrm{d}^{4}x\\
 & \equiv-\frac{1}{2}\int\mathfrak{S}^{\sigma\lambda\nu}\delta K_{\sigma\lambda\nu}\mathrm{d}^{4}x\\
 & \equiv-\frac{1}{2}\int S^{\sigma\lambda\nu}\delta K_{\sigma\lambda\nu}\mathrm{d}V,
\end{aligned}
\end{equation}
ensuring that since $K$ is anti-symmetric in its first two indices,
so are $S$ and $\mathfrak{S}$. Via our variation of $K$ in (\ref{eq:vary-K-in-grav-action}),
we then have the EOM
\begin{equation}
\begin{aligned}\kappa S^{\sigma\lambda\nu} & =K^{\mu\sigma}{}_{\mu}g^{\lambda\nu}+K^{\sigma\nu\lambda}-K^{\mu\lambda}{}_{\mu}g^{\sigma\nu}-K^{\lambda\nu\sigma},\end{aligned}
\end{equation}
which can be reversed to yield
\begin{equation}
\begin{aligned}K^{\sigma\lambda\nu} & =\frac{\kappa}{2}\left(S^{\sigma\lambda\nu}+S^{\nu\lambda\sigma}-S^{\nu\sigma\lambda}+g^{\lambda\nu}S^{\mu\sigma}{}_{\mu}-g^{\sigma\nu}S^{\mu\lambda}{}_{\mu}\right).\end{aligned}
\label{eq:K-in-terms-of-S}
\end{equation}
The spin tensor is instead usually written in terms of the torsion
\begin{equation}
\begin{aligned}\kappa S^{\sigma\lambda\nu} & =T^{\nu\sigma\lambda}-T^{\mu\sigma}{}_{\mu}g^{\lambda\nu}+T^{\mu\lambda}{}_{\mu}g^{\sigma\nu},\end{aligned}
\label{eq:2nd-einstein-cartan-eq}
\end{equation}
which is sometimes called the second Einstein-Cartan equation, and
can also be reversed to arrive at
\begin{equation}
\begin{aligned}T^{\nu\lambda\sigma} & =\frac{\kappa}{2}\left(2S^{\lambda\sigma\nu}-g^{\nu\lambda}S^{\mu\sigma}{}_{\mu}+g^{\nu\sigma}S^{\mu\lambda}{}_{\mu}\right).\end{aligned}
\label{eq:T-in-terms-of-S}
\end{equation}
Therefore the (con)torsion and spin tensor determine each other, and
if one vanishes so does the other: ``torsion due to spin does not
propagate.'' This is in contrast to curvature due to energy momentum,
which does propagate; i.e. the spacetime connection may not be flat
when $\overline{G}$ vanishes.

We now vary the metric with $K$ held constant. Recalling the expansion
of $R$ in (\ref{eq:expansion-of-R}), we have
\begin{equation}
\begin{aligned}2\kappa S_{\mathrm{G}}\left(g\right) & =\int\overline{R}\sqrt{g}\mathrm{d}^{4}x+2\int\overline{\nabla}_{\lambda}\left(g^{\mu\nu}K^{\lambda}{}_{\mu\nu}\right)\sqrt{g}\mathrm{d}^{4}x\\
 & \phantom{{}=}-\int g^{\mu\nu}\left(K^{\lambda}{}_{\mu\lambda}K^{\sigma}{}_{\nu\sigma}-K^{\lambda}{}_{\mu\sigma}K^{\sigma}{}_{\nu\lambda}\right)\sqrt{g}\mathrm{d}^{4}x.
\end{aligned}
\end{equation}
We already know the variation of the first integral yields the torsionless
Einstein tensor, while the variation of the second integral again
vanishes due to the divergence theorem, resulting in
\begin{equation}
\begin{aligned}2\kappa\delta S_{\mathrm{G}}\left(g\right) & =\int\overline{G}_{\mu\nu}\delta g^{\mu\nu}\mathrm{d}V\\
 & \phantom{{}=}-\int\left(K^{\lambda}{}_{\mu\lambda}K^{\sigma}{}_{\nu\sigma}-K^{\lambda}{}_{\mu\sigma}K^{\sigma}{}_{\nu\lambda}\right)\delta g^{\mu\nu}\mathrm{d}V\\
 & \phantom{{}=}+\int g^{\kappa\rho}\left(K^{\lambda}{}_{\kappa\lambda}K^{\sigma}{}_{\rho\sigma}-K^{\lambda}{}_{\kappa\sigma}K^{\sigma}{}_{\rho\lambda}\right)\frac{1}{2}g_{\mu\nu}\delta g^{\mu\nu}\mathrm{d}V\\
 & \equiv\int\left(\overline{G}_{\mu\nu}-\kappa\mathring{T}_{\mu\nu}\right)\delta g^{\mu\nu}\mathrm{d}V,
\end{aligned}
\end{equation}
where we define the \textbf{spin SEM tensor} by
\begin{equation}
\begin{aligned}\kappa\mathring{T}_{\mu\nu} & \equiv K^{\lambda}{}_{\mu\lambda}K^{\sigma}{}_{\nu\sigma}-K^{\lambda}{}_{\mu\sigma}K^{\sigma}{}_{\nu\lambda}-\frac{1}{2}g_{\mu\nu}g^{\kappa\rho}\left(K^{\lambda}{}_{\kappa\lambda}K^{\sigma}{}_{\rho\sigma}-K^{\lambda}{}_{\kappa\sigma}K^{\sigma}{}_{\rho\lambda}\right)\\
 & =K^{\lambda}{}_{\mu\lambda}K^{\sigma}{}_{\nu\sigma}-K^{\lambda}{}_{\mu\sigma}K^{\sigma}{}_{\nu\lambda}-\frac{1}{2}g_{\mu\nu}\left(K^{\lambda\rho}{}_{\lambda}K^{\sigma}{}_{\rho\sigma}-K^{\lambda\rho\sigma}K_{\sigma\rho\lambda}\right).
\end{aligned}
\label{eq:spin-SEM-tensor}
\end{equation}
The spin SEM tensor is manifestly symmetric, which is why we did not
have to symmetrize the variation above. We may substitute in (\ref{eq:K-in-terms-of-T})
or (\ref{eq:K-in-terms-of-S}) to find (even longer) expressions for
$\mathring{T}$ in terms of either the torsion or the spin tensor
(see \cite{Poplawski} p108). We also note that since it is quadratic
in $K$, which is linear in $S$, $\mathring{T}$ is also quadratic
in $S$, which means that it is unchanged by a change in the sign
of the spin tensor.

Our final EOM is then 
\begin{equation}
\begin{aligned}\overline{G}^{\mu\nu} & =\kappa\left(\overline{T}^{\mu\nu}+\mathring{T}^{\mu\nu}\right)\\
 & \equiv\kappa\mathring{\overline{T}}^{\mu\nu},
\end{aligned}
\label{eq:1st-einstein-cartan-eq}
\end{equation}
which is sometimes called the first Einstein-Cartan equation. $\mathring{T}$
in terms of the spin tensor thus may be viewed as the spin contribution
of matter to the geometric energy momentum $\overline{G}$. Note that
the Hilbert SEM tensor $\overline{T}$ remains symmetric, but is not
necessarily divergenceless. Only $\mathring{\overline{T}}$, the sum
of the Hilbert and spin SEM tensors, is divergenceless, since $\overline{G}$
is; we follow \cite{Poplawski} in calling it the \textbf{combined
SEM tensor}.

\subsection{\label{subsec:The-tetrad-SEM-tensor}The tetrad SEM tensor}

With our EOM complete under the variation of $g$ and $K$, we now
consider Einstein-Cartan-Sciama-Kibble  (ECSK) theory, in which torsion
is non-zero and we vary $e$ and $\omega$ in order to accommodate
actions such as that of Dirac theory. In this section we elect to
follow \cite{GockelerSchucker} and work in terms of differential
forms (for a treatment using tensors see \cite{Poplawski}). 

In order to make clear the types of each quantity, we temporarily
denote forms with an underbar. Thus the cotetrad is written $\underline{e}^{i}$,
four 1-forms labeled by $i$, whose components in its own frame are
$e_{j}^{i}=\delta_{j}^{i}$. Similarly, the Riemann curvature $\check{\underline{R}}=\mathrm{d}\check{\underline{\omega}}+\check{\underline{\omega}}\wedge\check{\underline{\omega}}$
is denoted $\check{R}_{ij}=R^{m}{}_{nij}=\underline{R}^{m}{}_{n}$,
which are the form components of a $so(3,1)$-valued 2-form, the Riemann
curvature tensor components, and a matrix of 2-forms respectively.
Since the variation of $\underline{e}$ is performed while holding
$\underline{\omega}^{i}{}_{j}$ constant, $\underline{R}^{m}{}_{n}$
is also constant under such a variation. All of these indices may
be raised and lowered using $\eta$, with the matrices $\underline{R}{}_{mn}$
then anti-symmetric. 

In preparation, we recall the component expression for the inner product
of two $k$-forms

\begin{equation}
\begin{aligned}\left\langle \underline{A},\underline{B}\right\rangle  & =\frac{1}{k!}A_{m_{1}\dots m_{k}}B^{m_{1}\dots m_{k}}\end{aligned}
\end{equation}
and the component expression for the exterior product of two 1-forms
\begin{equation}
\left(\underline{a}\wedge\underline{b}\right)_{mn}=a_{m}b_{n}-a_{n}b_{m}.
\end{equation}
We use the definition of the Hodge star of two $k$-forms
\begin{equation}
\underline{A}\wedge*\underline{B}=\left\langle \underline{A},\underline{B}\right\rangle \underline{\Omega},
\end{equation}
where $\underline{\Omega}=\mathrm{d}V$ is the volume element, which
in Lorentz indices may be written

\begin{align}
\begin{aligned}\left(*\underline{B}\right)_{m_{k+1}\dots m_{4}} & =\frac{1}{k!}B^{m_{1}\cdots m_{k}}\varepsilon_{m_{1}\cdots m_{4}}\\
\Rightarrow\left(*\underline{B}\right)_{m_{1}\dots m_{4-k}} & =\frac{-1}{k!}\varepsilon^{n_{1}\cdots n_{4}}B_{n_{1}\cdots n_{k}}\eta_{m_{1}n_{k+1}}\cdots\eta_{m_{n-k}n_{4}},
\end{aligned}
\label{eq:hodge-start-components}
\end{align}
where $\varepsilon$ is the completely anti-symmetric symbol. We also
recall that the generalized Kronecker delta
\begin{equation}
\delta_{m_{1}\cdots m_{k}}^{n_{1}\cdots n_{k}}\equiv\sum_{\pi}\textrm{sign}\left(\pi\right)\delta_{m_{1}}^{n_{\pi\left(1\right)}}\cdots\delta_{m_{k}}^{n_{\pi\left(k\right)}}
\end{equation}
gives the sign of the permutation of upper versus lower indices and
vanishes if they are not permutations or have a repeated index. We
can relate this to the permutation symbol via

\begin{equation}
\begin{aligned}\delta_{m_{1}\cdots m_{k}}^{n_{1}\cdots n_{k}} & =\frac{1}{\left(4-k\right)!}\varepsilon^{n_{1}\cdots n_{k}p_{k+1}\dots p_{4}}\varepsilon_{m_{1}\cdots m_{k}p_{k+1}\dots p_{4}}.\end{aligned}
\end{equation}

We first must write the gravitational Lagrangian in terms of forms.
Using the above, we find that
\begin{equation}
\begin{aligned}\left\langle \underline{R}{}_{mn},\underline{e}^{m}\wedge\underline{e}^{n}\right\rangle  & =\frac{1}{2}R_{mn}{}^{ij}\left(\underline{e}^{m}\wedge\underline{e}^{n}\right)_{ij}\\
 & =\frac{1}{2}R_{mn}{}^{ij}\left(e_{i}^{m}e_{j}^{n}-e_{j}^{m}e_{i}^{n}\right)\\
 & =\frac{1}{2}\left(R_{mn}{}^{mn}-R_{mn}{}^{nm}\right)\\
 & =R.
\end{aligned}
\end{equation}
We then use the definition of the Hodge star to obtain 
\begin{equation}
\begin{aligned}R\underline{\Omega} & =\left\langle \underline{R}_{mn},\underline{e}^{m}\wedge\underline{e}^{n}\right\rangle \underline{\Omega}\\
 & =\underline{R}_{mn}\wedge*\left(\underline{e}^{m}\wedge\underline{e}^{n}\right)\\
 & =\underline{R}_{mn}\wedge\frac{1}{2}\left(*\left(\underline{e}^{m}\wedge\underline{e}^{n}\right)\right)_{kl}\underline{e}^{k}\wedge\underline{e}^{l}\\
 & =\underline{R}_{mn}\wedge\frac{1}{4}\left(\underline{e}^{m}\wedge\underline{e}^{n}\right)^{ij}\varepsilon_{ijkl}\underline{e}^{k}\wedge\underline{e}^{l}\\
 & =\frac{1}{2}\underline{R}^{mn}\wedge\left(\underline{e}^{k}\wedge\underline{e}^{l}\right)\varepsilon_{mnkl},
\end{aligned}
\end{equation}
where in the last line we use the fact that $\left(\underline{e}^{m}\wedge\underline{e}^{n}\right)^{ij}$
vanishes except for $ij=mn$ or $nm$, in which cases we compensate
for the potential sign change due to re-lowering the indices by raising
the indices of $\underline{R}$.  

Varying the cotetrad in the gravitational action then yields
\begin{equation}
\begin{aligned}2\kappa\delta S_{\mathrm{G}}\left(\underline{e}\right) & =\delta\int R\underline{\Omega}\\
 & =\frac{1}{2}\int\left(\underline{R}^{mn}\wedge\delta\underline{e}^{i}\wedge\underline{e}^{j}+\underline{R}^{mn}\wedge\underline{e}^{i}\wedge\delta\underline{e}^{j}\right)\varepsilon_{mnij}\\
 & =\int\underline{R}^{mn}\wedge\delta\underline{e}^{i}\wedge\underline{e}^{j}\varepsilon_{mnij}\\
 & =\int\delta\underline{e}^{i}\wedge\underline{R}^{mn}\wedge\underline{e}^{j}\varepsilon_{imnj},
\end{aligned}
\end{equation}
where in the third line the sign change due to the reversed order
of the dual frames is cancelled by that of the permutation symbol,
and in the last line there is no sign change since $\underline{R}$
is a 2-form.

We now define the \textbf{tetrad SEM tensor} by 
\begin{equation}
\begin{aligned}\delta S_{\mathrm{M}}\left(\underline{e}\right) & \equiv-\int\overset{+}{\mathfrak{T}}_{\mu}{}^{i}\delta e_{i}^{\mu}\mathrm{d}^{4}x\\
 & \equiv\int\overset{+}{T}_{\mu}{}^{i}e_{i}^{\nu}e_{j}^{\mu}\delta e_{\nu}^{j}\mathrm{d}V\\
 & =\int\overset{+}{T}_{j}{}^{\nu}\delta e_{\nu}^{j}\mathrm{d}V\\
 & =\int\left\langle \overset{+}{\underline{T}}_{j},\delta\underline{e}^{j}\right\rangle \underline{\Omega}\\
 & =\int\delta\underline{e}^{j}\wedge*\overset{+}{\underline{T}}_{j},
\end{aligned}
\end{equation}
where $\overset{+}{\underline{T}}_{j}$ is a 1-form valued 1-form.
Requiring the variation of the entire action to vanish hence yields
the EOM
\begin{equation}
\begin{aligned}2\kappa\left(*\overset{+}{\underline{T}}_{i}\right) & =-\underline{R}^{mn}\wedge\underline{e}^{j}\varepsilon_{imnj}\\
\Rightarrow2\kappa\overset{+}{\underline{T}}_{n} & =-*\left(\underline{R}^{ij}\wedge\underline{e}^{m}\varepsilon_{nijm}\right)\\
 & =\frac{1}{2}R^{ij}{}_{kl}\varepsilon_{ijmn}\left(*\left(\underline{e}^{k}\wedge\underline{e}^{l}\wedge\underline{e}^{m}\right)\right)_{t}\underline{e}^{t}\\
 & =-\frac{1}{2}R^{ij}{}_{kl}\varepsilon_{ijmn}\varepsilon^{pqrs}\frac{1}{3!}\left(\underline{e}^{k}\wedge\underline{e}^{l}\wedge\underline{e}^{m}\right)_{pqr}\eta_{ts}\underline{e}^{t}\\
 & =-\frac{1}{2}R^{ij}{}_{kl}\varepsilon_{ijmn}\varepsilon^{klms}\eta_{ts}\underline{e}^{t}\\
 & =-\frac{1}{2}R^{ij}{}_{kl}\delta_{ijn}^{kls}\eta_{ts}\underline{e}^{t}\\
 & =-\left(R^{ij}{}_{ij}\delta_{n}^{s}-R^{ij}{}_{in}\delta_{j}^{s}+R^{ij}{}_{jn}\delta_{i}^{s}\right)\eta_{ts}\underline{e}^{t}\\
 & =-\left(R\delta_{n}^{s}-2R^{is}{}_{in}\right)\eta_{ts}\underline{e}^{t}\\
 & =2\left(R^{i}{}_{tin}-\frac{1}{2}R\eta_{tn}\right)\underline{e}^{t}\\
\Rightarrow\kappa\overset{+}{T}_{nt} & =G_{tn}\\
\Rightarrow G^{\mu\nu} & =\kappa\overset{+}{T}^{\nu\mu},
\end{aligned}
\label{eq:ECSK-EOM}
\end{equation}
where in the second line we use $**\overset{+}{\underline{T}}_{n}=\overset{+}{\underline{T}}_{n}$
since it is a 1-form, $G_{tn}$ is defined as the Einstein tensor
including torsion using Lorentz indices, and in the penultimate line
both sides act on $e_{t}$. 

For completeness, we now obtain the second Einstein-Cartan equation
(\ref{eq:2nd-einstein-cartan-eq}) by varying the spin connection.
We first note that the variation $\delta\check{\underline{\omega}}$
is a tensor-valued form since it is the difference between two connections,
and that we have
\begin{equation}
\begin{aligned}\mathrm{D}\delta\check{\underline{\omega}} & =\mathrm{d}\delta\check{\underline{\omega}}+\check{\underline{\omega}}\wedge\delta\check{\underline{\omega}}+\delta\check{\underline{\omega}}\wedge\check{\underline{\omega}}\\
\Rightarrow\delta\check{\underline{R}} & =\delta\left(\mathrm{d}\check{\underline{\omega}}+\check{\underline{\omega}}\wedge\check{\underline{\omega}}\right)\\
 & =\mathrm{d}\delta\check{\underline{\omega}}+\delta\check{\underline{\omega}}\wedge\check{\underline{\omega}}+\check{\underline{\omega}}\wedge\delta\check{\underline{\omega}}\\
 & =\mathrm{D}\delta\check{\underline{\omega}},
\end{aligned}
\end{equation}
where we recall the expressions for the exterior covariant derivative
of such forms (see e.g. \cite{Marsh-book} Sections 9.2.3 and 3.3.5).
For the gravitational action, we therefore have
\begin{equation}
\begin{aligned}2\kappa\delta S_{\mathrm{G}}\left(\check{\underline{\omega}}\right) & =\delta\int R\underline{\Omega}\\
 & =\frac{1}{2}\int\delta\underline{R}^{mn}\wedge\underline{e}^{i}\wedge\underline{e}^{j}\varepsilon_{mnij}\\
 & =\frac{1}{2}\int\left(\mathrm{D}\delta\underline{\omega}\right)^{mn}\wedge\underline{e}^{i}\wedge\underline{e}^{j}\varepsilon_{mnij}.
\end{aligned}
\end{equation}
Now, the quantity $\delta\underline{\omega}^{mn}\wedge\underline{e}^{i}\wedge\underline{e}^{j}\varepsilon_{mnij}$
is a scalar-valued 3-form, so the exterior covariant derivative is
the same as the exterior derivative; furthermore, we recall (see e.g.
\cite{Marsh-book} Section 9.2.2) that $\mathrm{D}$ is a graded derivation
with respect to the exterior product of anti-symmetric tensor-valued
forms. Therefore we may use Stokes' theorem and the assumption that
$\delta\check{\omega}$ vanishes on the boundary to write
\begin{equation}
\begin{aligned}0 & =\int\mathrm{d}\left(\delta\underline{\omega}^{mn}\wedge\underline{e}^{i}\wedge\underline{e}^{j}\varepsilon_{mnij}\right)\\
 & =\int\mathrm{D}\left(\delta\underline{\omega}^{mn}\wedge\underline{e}^{i}\wedge\underline{e}^{j}\varepsilon_{mnij}\right)\\
 & =\int\left(\mathrm{D}\delta\underline{\omega}\right)^{mn}\wedge\underline{e}^{i}\wedge\underline{e}^{j}\varepsilon_{mnij}-2\int\delta\underline{\omega}^{mn}\wedge\left(\mathrm{D}\underline{e}\right)^{i}\wedge\underline{e}^{j}\varepsilon_{mnij},
\end{aligned}
\end{equation}
where again the sign change in the order of the dual frame terms is
cancelled by that of the permutation symbol. Recalling (see e.g. \cite{Marsh-book}
Section 9.2.4) the definition of torsion in terms of the dual frame
as a vector-valued 1-form, we then have
\begin{equation}
\begin{aligned}2\kappa\delta S_{\mathrm{G}}\left(\check{\underline{\omega}}\right) & =\int\delta\underline{\omega}^{mn}\wedge\left(\mathrm{D}\underline{e}\right)^{i}\wedge\underline{e}^{j}\varepsilon_{mnij}\\
 & =\int\delta\underline{\omega}^{mn}\wedge\underline{T}^{i}\wedge\underline{e}^{j}\varepsilon_{mnij},
\end{aligned}
\end{equation}
where $\underline{T}^{i}$ is the torsion as a vector-valued 2-form. 

Proceeding to the matter action, we have
\begin{equation}
\begin{aligned}\delta S_{\mathrm{M}}\left(\check{\underline{\omega}}\right) & =-\frac{1}{2}\int\mathfrak{S}_{i}{}^{j\mu}\delta\omega^{i}{}_{j\mu}\mathrm{d}^{4}x\\
 & =-\frac{1}{2}\int S_{ij}{}^{k}\delta\omega^{ij}{}_{k}\mathrm{d}V\\
 & =-\frac{1}{2}\int\left\langle \underline{S}_{ij},\delta\underline{\omega}^{ij}\right\rangle \Omega\\
 & =-\frac{1}{2}\int\delta\underline{\omega}^{ij}\wedge*\underline{S}_{ij},
\end{aligned}
\end{equation}
where $\underline{S}_{ij}$ is a 2-form valued 1-form. Requiring the
variation of the entire action to vanish hence yields the EOM
\begin{equation}
\begin{aligned}\kappa\left(*\underline{S}_{ij}\right) & =\underline{T}^{k}\wedge\underline{e}^{l}\varepsilon_{ijkl}\\
\Rightarrow\kappa\underline{S}_{ij} & =*\left(\underline{T}^{k}\wedge\underline{e}^{l}\varepsilon_{ijkl}\right)\\
 & =\frac{1}{2}T^{k}{}_{mn}\varepsilon_{ijkl}\left(*\left(\underline{e}^{m}\wedge\underline{e}^{n}\wedge\underline{e}^{l}\right)\right)_{t}\underline{e}^{t}\\
 & =-\frac{1}{2}T^{k}{}_{mn}\varepsilon_{ijkl}\varepsilon^{pqrs}\frac{1}{3!}\left(\underline{e}^{m}\wedge\underline{e}^{n}\wedge\underline{e}^{l}\right)_{pqr}\eta_{ts}\underline{e}^{t}\\
 & =-\frac{1}{2}T^{k}{}_{mn}\varepsilon_{ijkl}\varepsilon^{mnls}\eta_{ts}\underline{e}^{t}\\
 & =\frac{1}{2}T^{k}{}_{mn}\delta_{ijk}^{mns}\eta_{ts}\underline{e}^{t}\\
 & =\left(T^{k}{}_{ij}\delta_{k}^{s}-T^{k}{}_{ik}\delta_{j}^{s}+T^{k}{}_{jk}\delta_{i}^{s}\right)\eta_{ts}\underline{e}^{t}\\
 & =\left(T_{tij}-T^{k}{}_{ik}\eta_{tj}+T^{k}{}_{jk}\eta_{ti}\right)\underline{e}^{t}\\
\Rightarrow\kappa S_{ijt} & =T_{tij}-T^{k}{}_{ik}\eta_{tj}+T^{k}{}_{jk}\eta_{ti},
\end{aligned}
\end{equation}
where again in the second line we use $**\underline{S}_{ij}=\underline{S}_{ij}$
since it is a 1-form, and again in the last line both sides act on
$\underline{e}_{t}$. This completes the derivation of the second
Einstein-Cartan equation via variation of the spin connection.

\subsection{\label{subsec:The-Belinfante-Rosenfeld-relation}The Belinfante-Rosenfeld
relation}

We associate the combined SEM tensor (\ref{eq:1st-einstein-cartan-eq})
with total energy momentum, and therefore would like to obtain it
in theories such as Dirac theory which are expressed in terms of $e$
and $\omega$. The Belinfante-Rosenfeld relation, derived in this
section, accomplishes this by expressing $\overline{T}$ in terms
of $\overset{+}{T}$ and $S$. 

Using the ECSK EOM (\ref{eq:ECSK-EOM}) and the curvature expansion
(\ref{eq:expansion-of-R}), we have
\begin{align}
\begin{aligned}G_{\nu\mu} & =\kappa\overset{+}{T}_{\mu\nu}\\
 & =R_{\nu\mu}-\frac{1}{2}g_{\nu\mu}R\\
 & =\overline{G}_{\mu\nu}+\overline{\nabla}_{\lambda}K^{\lambda}{}_{\nu\mu}-\overline{\nabla}_{\mu}K^{\lambda}{}_{\nu\lambda}+K^{\lambda}{}_{\sigma\lambda}K^{\sigma}{}_{\nu\mu}-K^{\lambda}{}_{\sigma\mu}K^{\sigma}{}_{\nu\lambda}\\
 & \phantom{{}=}-g_{\mu\nu}\overline{\nabla}_{\lambda}\left(g^{\kappa\rho}K^{\lambda}{}_{\kappa\rho}\right)-\frac{1}{2}g_{\mu\nu}\left(K^{\lambda}{}_{\sigma\lambda}K^{\sigma\kappa}{}_{\kappa}-K^{\lambda}{}_{\sigma\kappa}K^{\sigma\kappa}{}_{\lambda}\right).
\end{aligned}
\end{align}
Using the EOM (\ref{eq:1st-einstein-cartan-eq}) and the spin SEM
tensor definition (\ref{eq:spin-SEM-tensor}), we thus have
\begin{align}
\begin{aligned}\Rightarrow\kappa\overline{T}_{\mu\nu} & =\overline{G}_{\mu\nu}-\kappa\mathring{T}_{\mu\nu}\\
 & =\kappa\overset{+}{T}_{\mu\nu}-\overline{\nabla}_{\lambda}K^{\lambda}{}_{\nu\mu}+\overline{\nabla}_{\mu}K^{\lambda}{}_{\nu\lambda}-K^{\lambda}{}_{\sigma\lambda}K^{\sigma}{}_{\nu\mu}+K^{\lambda}{}_{\sigma\mu}K^{\sigma}{}_{\nu\lambda}\\
 & \phantom{{}=}+g_{\mu\nu}\overline{\nabla}_{\lambda}\left(g^{\kappa\rho}K^{\lambda}{}_{\kappa\rho}\right)-\frac{1}{2}g_{\mu\nu}\left(K^{\lambda\sigma}{}_{\lambda}K^{\kappa}{}_{\sigma\kappa}-K^{\lambda\sigma\kappa}K_{\kappa\sigma\lambda}\right)\\
 & \phantom{{}=}-K^{\lambda}{}_{\mu\lambda}K^{\sigma}{}_{\nu\sigma}+K^{\lambda}{}_{\mu\sigma}K^{\sigma}{}_{\nu\lambda}+\frac{1}{2}g_{\mu\nu}\left(K^{\lambda\rho}{}_{\lambda}K^{\sigma}{}_{\rho\sigma}-K^{\lambda\rho\sigma}K_{\sigma\rho\lambda}\right)\\
 & =\kappa\overset{+}{T}_{\mu\nu}-\left(g_{\mu\nu}\overline{\nabla}_{\lambda}K^{\kappa\lambda}{}_{\kappa}+\overline{\nabla}_{\lambda}K^{\lambda}{}_{\nu\mu}-K^{\sigma}{}_{\nu\lambda}K^{\lambda}{}_{\sigma\mu}-K^{\sigma}{}_{\mu\lambda}K^{\lambda}{}_{\nu\sigma}\right)\\
 & \phantom{{}=}+\left(\overline{\nabla}_{\mu}K^{\lambda}{}_{\nu\lambda}-K^{\sigma}{}_{\nu\mu}K^{\lambda}{}_{\sigma\lambda}\right)-K^{\lambda}{}_{\mu\lambda}K^{\sigma}{}_{\nu\sigma}\\
 & =\kappa\overset{+}{T}_{\mu\nu}-\overline{\nabla}_{\lambda}\left(K^{\kappa\lambda}{}_{\kappa}g_{\mu\nu}+K^{\lambda}{}_{\nu\mu}\right)-K^{\lambda}{}_{\mu\lambda}K^{\sigma}{}_{\nu\sigma}\\
 & \phantom{{}=}+\overline{\nabla}_{\mu}K^{\lambda}{}_{\nu\lambda}-K^{\lambda}{}_{\sigma\mu}K^{\sigma}{}_{\nu\lambda}+K^{\sigma}{}_{\mu\lambda}K^{\lambda}{}_{\nu\sigma}-K^{\sigma}{}_{\nu\mu}K^{\lambda}{}_{\sigma\lambda}\\
 & =\kappa\overset{+}{T}_{\mu\nu}-\left(\nabla_{\lambda}-K^{\rho}{}_{\lambda\rho}\right)\left(K^{\kappa\lambda}{}_{\kappa}g_{\mu\nu}+K^{\lambda}{}_{\nu\mu}-K^{\kappa}{}_{\nu\kappa}\delta_{\mu}^{\lambda}\right)\\
\Rightarrow\kappa\overline{T}^{\mu\nu} & =\kappa\overset{+}{T}^{\mu\nu}-\left(\nabla_{\lambda}-K^{\rho}{}_{\lambda\rho}\right)\left(K^{\kappa\lambda}{}_{\kappa}g^{\mu\nu}+K^{\lambda\nu\mu}-K^{\kappa\nu}{}_{\kappa}g^{\mu\lambda}\right).
\end{aligned}
\end{align}

To obtain this expression in terms of the spin tensor we could use
(\ref{eq:K-in-terms-of-S}), but to ease calculations we instead use
(\ref{eq:K-in-terms-of-T}) to get
\begin{align}
\begin{aligned}K_{\lambda\mu\lambda} & =\frac{1}{2}\left(T_{\lambda\lambda\mu}-T_{\lambda\mu\lambda}\right)\\
\Rightarrow K^{\lambda}{}_{\mu\lambda} & =-T^{\lambda}{}_{\mu\lambda}\\
\Rightarrow\kappa\overline{T}^{\mu\nu} & =\kappa\overset{+}{T}^{\mu\nu}-\left(\nabla_{\lambda}+T^{\rho}{}_{\lambda\rho}\right)A^{\lambda\nu\mu},\\
A^{\lambda\nu\mu} & \equiv K^{\kappa\lambda}{}_{\kappa}g^{\mu\nu}+K^{\lambda\nu\mu}-K^{\kappa\nu}{}_{\kappa}g^{\mu\lambda}\\
 & =-T^{\kappa\lambda}{}_{\kappa}g^{\mu\nu}+T^{\kappa\nu}{}_{\kappa}g^{\mu\lambda}+\frac{1}{2}\left(T^{\nu\lambda\mu}+T^{\mu\lambda\nu}-T^{\lambda\nu\mu}\right),
\end{aligned}
\end{align}
then use (\ref{eq:2nd-einstein-cartan-eq}) to get
\begin{equation}
\begin{aligned}\kappa S^{\lambda\nu\mu} & =T^{\mu\lambda\nu}-T^{\kappa\lambda}{}_{\kappa}g^{\mu\nu}+T^{\kappa\nu}{}_{\kappa}g^{\mu\lambda}\\
\Rightarrow A^{\lambda\nu\mu} & =\kappa S^{\lambda\nu\mu}+\frac{1}{2}\left(T^{\nu\lambda\mu}-T^{\mu\lambda\nu}-T^{\lambda\nu\mu}\right),
\end{aligned}
\end{equation}
so that using (\ref{eq:T-in-terms-of-S}) we have
\begin{equation}
\begin{aligned}A^{\lambda\nu\mu} & =\kappa S^{\lambda\nu\mu}+\frac{\kappa}{4}\left(2S^{\lambda\mu\nu}-g^{\nu\lambda}S^{\kappa\mu}{}_{\kappa}+g^{\nu\mu}S^{\kappa\lambda}{}_{\kappa}\right)\\
 & \phantom{{}=}-\frac{\kappa}{4}\left(2S^{\lambda\nu\mu}-g^{\mu\lambda}S^{\kappa\nu}{}_{\kappa}+g^{\nu\mu}S^{\kappa\lambda}{}_{\kappa}\right)\\
 & \phantom{{}=}-\frac{\kappa}{4}\left(2S^{\nu\mu\lambda}-g^{\nu\lambda}S^{\kappa\mu}{}_{\kappa}+g^{\lambda\mu}S^{\kappa\nu}{}_{\kappa}\right)\\
 & =\frac{\kappa}{2}\left(S^{\lambda\mu\nu}+S^{\lambda\nu\mu}-S^{\nu\mu\lambda}\right),
\end{aligned}
\end{equation}
yielding our final relation
\begin{equation}
\begin{aligned}\Rightarrow\overline{T}{}^{\mu\nu} & =\overset{+}{T}{}^{\mu\nu}-\frac{1}{2}\left(\nabla_{\lambda}+T^{\rho}{}_{\lambda\rho}\right)\left(S^{\lambda\mu\nu}+S^{\lambda\nu\mu}-S^{\nu\mu\lambda}\right).\end{aligned}
\label{eq:belinfante-rosenfeld-relation}
\end{equation}
This is the Belinfante-Rosenfeld relation, our desired expression
for the Hilbert SEM tensor in terms of the tetrad SEM tensor and the
spin tensor. For completeness, we include a lengthier alternative
derivation in Appendix \ref{sec:alt-Belinfante-Rosenfeld-derivation}
which is obtained by matching variations, detailing that of \cite{Poplawski}
with our conventions.

In flat torsionless spacetime, theories such as Dirac theory have
actions in which the derivative of the field appears in the same place
where the tetrad appears in curved spacetime; therefore the tetrad
SEM tensor is equal to the canonical SEM tensor derived from the Noether
current under a constant displacement (the negative sign is not present
when using the mostly minuses signature). Furthermore, the Noether
current associated with Lorentz transformations is identical to the
spin tensor, in which context it is called the canonical spin tensor
(which does not vanish with torsion as does the spin tensor). In this
context the Belinfante-Rosenfeld relation is characterized as ``improving''
the canonical SEM tensor, i.e. transforming it into the Hilbert SEM
tensor, which has the desired properties of being symmetric and divergenceless.

\subsection{\label{sec:Summary}Summary}

Summarizing our conventions and results, we have for the connections
\begin{equation}
\begin{aligned}\Gamma^{\lambda}{}_{\sigma\mu} & =\mathrm{d}x^{\lambda}\left(\nabla_{\mu}\partial_{\sigma}\right),\\
\omega^{j}{}_{i\mu} & =e^{j}\left(\nabla_{\mu}e_{i}\right),
\end{aligned}
\end{equation}
for the torsion
\begin{align}
\begin{aligned}T^{i}{}_{jk} & =\omega^{i}{}_{kj}-\omega^{i}{}_{jk}-[e_{j},e_{k}]^{i},\\
T^{\lambda}{}_{\mu\nu} & =e_{i}^{\lambda}\left(e_{\nu}^{j}\omega^{i}{}_{j\mu}-e_{\mu}^{j}\omega^{i}{}_{j\nu}+\partial_{\mu}e_{\nu}^{i}-\partial_{\nu}e_{\mu}^{i}\right),
\end{aligned}
\end{align}
and for the contorsion
\begin{align}
\begin{aligned}K_{\lambda\mu\nu} & =\frac{1}{2}\left(T_{\mu\lambda\nu}+T_{\nu\lambda\mu}-T_{\lambda\mu\nu}\right),\\
\Gamma^{\lambda}{}_{\mu\nu} & =\overline{\Gamma}^{\lambda}{}_{\mu\nu}+K^{\lambda}{}_{\mu\nu}.
\end{aligned}
\end{align}

Varying the parallel transport in the gravitational action via the
metric and contorsion, we have 
\begin{equation}
\begin{aligned}\delta S_{\mathrm{G}}\left(g\right)+\delta S_{\mathrm{G}}\left(K\right) & =-\frac{1}{2\kappa}\int\left(\overline{G}^{\mu\nu}-\kappa\mathring{T}^{\mu\nu}\right)\delta g_{\mu\nu}\mathrm{d}V\\
 & \phantom{{}=}+\frac{1}{2\kappa}\int\left(K^{\mu\sigma}{}_{\mu}g^{\lambda\nu}+K^{\sigma\nu\lambda}-K^{\mu\lambda}{}_{\mu}g^{\sigma\nu}-K^{\lambda\nu\sigma}\right)\delta K_{\sigma\lambda\nu}\mathrm{d}V,
\end{aligned}
\end{equation}
where the torsionless Einstein tensor is 
\begin{equation}
\overline{G}_{\mu\nu}=\overline{R}^{\mu\nu}-\frac{1}{2}\overline{R}g^{\mu\nu}
\end{equation}
and the spin SEM tensor is
\begin{equation}
\begin{aligned}\kappa\mathring{T}^{\mu\nu} & =K^{\lambda\mu}{}_{\lambda}K^{\sigma\nu}{}_{\sigma}-K^{\lambda\mu}{}_{\sigma}K^{\sigma\nu}{}_{\lambda}-\frac{1}{2}g^{\mu\nu}\left(K^{\lambda\rho}{}_{\lambda}K^{\sigma}{}_{\rho\sigma}-K^{\lambda\rho\sigma}K_{\sigma\rho\lambda}\right).\end{aligned}
\end{equation}
The variation of the matter action may be written 
\begin{equation}
\begin{aligned}\delta S_{\mathrm{M}}\left(g\right)+\delta S_{\mathrm{M}}\left(K\right) & =\frac{1}{2}\int\overline{T}^{\mu\nu}\delta g_{\mu\nu}\mathrm{d}V-\frac{1}{2}\int S^{\sigma\lambda\nu}\delta K_{\sigma\lambda\nu}\mathrm{d}V,\end{aligned}
\end{equation}
where the Hilbert SEM tensor is
\begin{equation}
\begin{aligned}\overline{T}^{\mu\nu} & =\frac{1}{\sqrt{g}}\left(\frac{\partial\mathfrak{L}_{\mathrm{M}}}{\partial g_{\mu\nu}}+\frac{\partial\mathfrak{L}_{\mathrm{M}}}{\partial g_{\nu\mu}}\right)\end{aligned}
\end{equation}
and the spin tensor is 
\begin{equation}
S^{\sigma\lambda\nu}=-\frac{1}{\sqrt{g}}\left(\frac{\partial\mathfrak{L}_{\mathrm{M}}}{\partial K_{\sigma\lambda\nu}}-\frac{\partial\mathfrak{L}_{\mathrm{M}}}{\partial K_{\lambda\sigma\nu}}\right).
\end{equation}
Our EOM are then 
\begin{equation}
\begin{aligned}\overline{G}^{\mu\nu} & =\kappa\left(\overline{T}^{\mu\nu}+\mathring{T}^{\mu\nu}\right),\\
K^{\mu\sigma}{}_{\mu}g^{\lambda\nu}+K^{\sigma\nu\lambda}-K^{\mu\lambda}{}_{\mu}g^{\sigma\nu}-K^{\lambda\nu\sigma} & =-\kappa S^{\sigma\lambda\nu},
\end{aligned}
\end{equation}
which can be written
\begin{equation}
\begin{aligned}\overline{G}^{\mu\nu} & =\kappa\mathring{\overline{T}}^{\mu\nu},\\
T^{\nu\sigma\lambda}-T^{\mu\sigma}{}_{\mu}g^{\lambda\nu}+T^{\mu\lambda}{}_{\mu}g^{\sigma\nu} & =\kappa S^{\sigma\lambda\nu},
\end{aligned}
\end{equation}
where the combined SEM tensor is 
\begin{equation}
\mathring{\overline{T}}^{\mu\nu}=\overline{T}^{\mu\nu}+\mathring{T}^{\mu\nu}.
\end{equation}
If the matter action does not depend on $K$ then $S=0$, and less
obviously $K=0$, so that the EOM revert to the Einstein field equations
$\overline{G}=\kappa\overline{T}$.

Now varying the parallel transport in the gravitational action via
the cotetrad and spin connection, we have 
\begin{equation}
\begin{aligned}\delta S_{\mathrm{G}}\left(\underline{e}\right)+\delta S_{\mathrm{G}}\left(\underline{\omega}\right) & =\frac{1}{2\kappa}\int\delta\underline{e}^{i}\wedge\underline{R}^{mn}\wedge\underline{e}^{j}\varepsilon_{imnj}+\frac{1}{2\kappa}\int\delta\underline{\omega}^{mn}\wedge\underline{T}^{i}\wedge\underline{e}^{j}\varepsilon_{mnij},\end{aligned}
\end{equation}
where $\underline{R}^{mn}$ is the Riemann curvature tensor as a matrix
of 2-forms and $\underline{T}^{i}$ is the torsion as a vector-valued
2-form. The variation of the matter action may be written 
\begin{equation}
\begin{aligned}\delta S_{\mathrm{M}}\left(\underline{e}\right)+\delta S_{\mathrm{M}}\left(\underline{\omega}\right) & =\int\delta\underline{e}^{j}\wedge*\overset{+}{\underline{T}}_{j}-\frac{1}{2}\int\delta\underline{\omega}^{ij}\wedge*\underline{S}_{ij},\end{aligned}
\end{equation}
where the tetrad SEM tensor is
\begin{equation}
\begin{aligned}\overset{+}{T}_{\mu}{}^{i} & =-\frac{1}{\sqrt{g}}\frac{\partial\mathfrak{L}_{\mathrm{M}}}{\partial e_{i}^{\mu}}\end{aligned}
,
\end{equation}
which may be written as a vector valued 1-form $\overset{+}{\underline{T}}^{i}$,
and $\underline{S}_{ij}$ is the spin tensor as a 2-form valued 1-form.
The resulting EOM may be written 
\begin{equation}
\begin{aligned}G_{ij} & =\kappa\overset{+}{T}_{ji},\\
T_{tij}-T^{k}{}_{ik}\eta_{tj}+T^{k}{}_{jk}\eta_{ti} & =\kappa S_{ijt},
\end{aligned}
\end{equation}
where 
\begin{equation}
G_{ij}=R_{ij}-\frac{1}{2}R\eta_{ij}
\end{equation}
is the Einstein tensor including torsion. The second EOM is identical
to that previously found by varying the contorsion, as expected from
(\ref{eq:var-K-same-as-var-omega}). 

The first EOM $G_{ij}=\kappa\overset{+}{T}_{ji}$ may be expanded
by writing $G$ in terms of $\overline{G}$ and $K$, the former of
which may be written in terms of $\overline{T}$ and $K$, by which
we obtain the Belinfante-Rosenfeld relation
\begin{equation}
\begin{aligned}\kappa\overline{T}_{\mu\nu} & =\kappa\overset{+}{T}_{\mu\nu}-\left(\nabla_{\lambda}-K^{\rho}{}_{\lambda\rho}\right)\left(K^{\kappa\lambda}{}_{\kappa}g^{\mu\nu}+K^{\lambda\nu\mu}-K^{\kappa\nu}{}_{\kappa}g^{\mu\lambda}\right),\end{aligned}
\end{equation}
which can be written in terms of the spin tensor as
\begin{equation}
\begin{aligned}\overline{T}{}^{\mu\nu} & =\overset{+}{T}{}^{\mu\nu}-\frac{1}{2}\left(\nabla_{\lambda}+T^{\rho}{}_{\lambda\rho}\right)\left(S^{\lambda\mu\nu}+S^{\lambda\nu\mu}-S^{\nu\mu\lambda}\right).\end{aligned}
\end{equation}
When considering an action defined in terms of $e$ and $\omega$,
such as that of Dirac theory, this relation allows us to obtain the
Hilbert SEM tensor $\overline{T}$ from calculations of $\overset{+}{T}$
and $S$. The spin SEM tensor $\mathring{T}$ may also be obtained
from $S$ via (\ref{eq:K-in-terms-of-S}) and (\ref{eq:spin-SEM-tensor}).
Summing these yields the combined SEM tensor $\mathring{\overline{T}}$,
which being proportional to the torsionless Einstein tensor $\overline{G}$,
allows us to determine energy and momentum as defined by geodesic
accelerations for such an action.

\appendix

\section{\label{sec:alt-Belinfante-Rosenfeld-derivation}The Belinfante-Rosenfeld
relation via matched variations}

Here we derive the Belinfante-Rosenfeld relation in a way which details
that of \cite{Poplawski} with our conventions, starting with the
observation that the variation of the parallel transport or connection
is equivalent to the independent variations of $e$ and $\omega$,
which induce variations of $g$ and the torsion $T$. When varying
$\omega$, we recall (\ref{eq:var-K-same-as-var-omega}) to obtain
\begin{equation}
\begin{aligned}\delta S_{\mathrm{M}}\left(\omega\right) & =\int\frac{\partial\mathfrak{L}_{\mathrm{M}}}{\partial\omega^{i}{}_{j\mu}}\delta\omega^{i}{}_{j\mu}\mathrm{d}^{4}x\\
 & =\int\frac{\partial\mathfrak{L}_{\mathrm{M}}}{\partial K^{i}{}_{j\mu}}\frac{\partial K^{i}{}_{j\mu}}{\partial\omega^{i}{}_{j\mu}}\delta\omega^{i}{}_{j\mu}\mathrm{d}^{4}x\\
 & =-\frac{1}{2}\int\mathfrak{S}_{i}{}^{j\mu}\delta\omega^{i}{}_{j\mu}\mathrm{d}^{4}x,
\end{aligned}
\end{equation}
since we can convert indices inside $\delta K^{\lambda}{}_{\sigma\mu}$
with the tetrad held constant, and the anti-symmetry of the integrand
is already present. 

Now, recall that varying $K$ while holding $g$ constant is equivalent
to varying the torsion $T$ while holding $g$ constant; we therefore
may express the variational derivative of the matter action with respect
to the torsion in terms of the spin tensor:
\begin{equation}
\begin{aligned}\delta S_{\mathrm{M}}\left(K\right) & =-\frac{1}{2}\int\mathfrak{S}^{\lambda\mu\nu}\delta K_{\lambda\mu\nu}\mathrm{d}^{4}x\\
 & =-\frac{1}{4}\int\mathfrak{S}^{\lambda\mu\nu}\delta\left(T_{\mu\lambda\nu}+T_{\nu\lambda\mu}-T_{\lambda\mu\nu}\right)\mathrm{d}^{4}x\\
 & =-\frac{1}{4}\int\left(\mathfrak{S}^{\mu\lambda\nu}+\mathfrak{S}^{\mu\nu\lambda}-\mathfrak{S}^{\lambda\mu\nu}\right)\delta T_{\lambda\mu\nu}\mathrm{d}^{4}x\\
\Rightarrow\delta S_{\mathrm{M}}\left(T\right) & =-\frac{1}{8}\int\left(\mathfrak{S}^{\mu\lambda\nu}+\mathfrak{S}^{\mu\nu\lambda}-\mathfrak{S}^{\lambda\mu\nu}-\mathfrak{S}^{\nu\lambda\mu}-\mathfrak{S}^{\nu\mu\lambda}+\mathfrak{S}^{\lambda\nu\mu}\right)\delta T_{\lambda\mu\nu}\mathrm{d}^{4}x\\
 & =-\frac{1}{4}\int\left(\mathfrak{S}^{\mu\lambda\nu}+\mathfrak{S}^{\mu\nu\lambda}-\mathfrak{S}^{\nu\lambda\mu}\right)\delta T_{\lambda\mu\nu}\mathrm{d}^{4}x\\
 & \equiv-\int\mathfrak{N}^{\lambda\mu\nu}\delta T_{\lambda\mu\nu}\mathrm{d}^{4}x\\
\Rightarrow\mathfrak{\mathfrak{N}}^{\lambda\mu\nu} & =\frac{1}{4}\left(\mathfrak{S}^{\mu\lambda\nu}+\mathfrak{S}^{\mu\nu\lambda}-\mathfrak{S}^{\nu\lambda\mu}\right),\\
\mathfrak{S}^{\mu\nu\lambda} & =2\left(\mathfrak{\mathfrak{N}}^{\nu\mu\lambda}-\mathfrak{\mathfrak{N}}^{\mu\nu\lambda}\right).
\end{aligned}
\end{equation}
In the fourth line above we have relabeled dummy indices and then
anti-symmetrized in order to reflect a variation of the torsion 2-form
underlying the torsion tensor, which ensures that $\mathfrak{\mathfrak{N}}^{\lambda\mu\nu}$
is anti-symmetric in its last two indices. 

Our variations and induced variations are then 
\begin{equation}
\begin{aligned}\delta S_{\mathrm{M}}\left(\Gamma\right)=\delta S_{\mathrm{M}}\left(e\right)+\delta S_{\mathrm{M}}\left(\omega\right) & =-\int\overset{+}{\mathfrak{T}}_{\mu}{}^{i}\delta e_{i}^{\mu}\mathrm{d}^{4}x-\frac{1}{2}\int\mathfrak{S}_{i}{}^{j\mu}\delta\omega^{i}{}_{j\mu}\mathrm{d}^{4}x\\
 & =\frac{1}{2}\int\overline{\mathfrak{T}}^{\mu\nu}\delta g_{\mu\nu}\mathrm{d}^{4}x-\int\mathfrak{\mathfrak{N}}_{\lambda}{}^{\mu\nu}\delta T^{\lambda}{}_{\mu\nu}\mathrm{d}^{4}x.
\end{aligned}
\label{eq:variations-and-induced-variations}
\end{equation}
Now, the first term in the second line of (\ref{eq:variations-and-induced-variations})
can be written
\begin{equation}
\begin{aligned}\frac{1}{2}\int\overline{\mathfrak{T}}^{\mu\nu}\delta g_{\mu\nu}\mathrm{d}^{4}x & =-\frac{1}{2}\int\overline{\mathfrak{T}}_{\mu\nu}\delta g^{\mu\nu}\mathrm{d}^{4}x\\
 & =-\frac{1}{2}\int\overline{\mathfrak{T}}_{\mu\nu}\delta\left(e_{i}^{\mu}e_{j}^{\nu}\eta^{ij}\right)\mathrm{d}^{4}x\\
 & =-\frac{1}{2}\int\overline{\mathfrak{T}}_{\mu\nu}\eta^{ij}\left(e_{j}^{\nu}\delta e_{i}^{\mu}+e_{i}^{\mu}\delta e_{j}^{\nu}\right)\mathrm{d}^{4}x\\
 & =-\int\overline{\mathfrak{T}}_{\mu\nu}e^{i\nu}\delta e_{i}^{\mu}\mathrm{d}^{4}x\\
 & =-\int\overline{\mathfrak{T}}_{\mu}{}^{i}\delta e_{i}^{\mu}\mathrm{d}^{4}x,
\end{aligned}
\end{equation}
where we have used the symmetry of both $\eta$ and $\mathfrak{T}$.
Note that this is consistent with the first line of (\ref{eq:variations-and-induced-variations}),
since the variation of the torsion also includes terms in the variation
of the tetrad. Our calculation thus comes down to writing the torsion
variation in terms of variations of $e$ and $\omega$, so that we
may equate them. With some perseverance, we find that
\begin{equation}
\begin{aligned}\int\mathfrak{\mathfrak{N}}_{\lambda}{}^{\mu\nu}\delta T^{\lambda}{}_{\mu\nu}\mathrm{d}^{4}x & =\int\mathfrak{\mathfrak{N}}_{\lambda}{}^{\mu\nu}\delta\left(e_{i}^{\lambda}e_{\nu}^{k}\omega^{i}{}_{k\mu}-e_{i}^{\lambda}e_{\mu}^{j}\omega^{i}{}_{j\nu}+e_{i}^{\lambda}\partial_{\mu}e_{\nu}^{i}-e_{i}^{\lambda}\partial_{\nu}e_{\mu}^{i}\right)\mathrm{d}^{4}x\\
 & =2\int\mathfrak{\mathfrak{N}}_{\lambda}{}^{\mu\nu}\delta\left(e_{i}^{\lambda}e_{\nu}^{j}\omega^{i}{}_{j\mu}+e_{i}^{\lambda}\partial_{\mu}e_{\nu}^{i}\right)\mathrm{d}^{4}x\\
 & =2\int\left(\mathfrak{\mathfrak{N}}_{\lambda}{}^{\mu\nu}\left(\delta e_{i}^{\lambda}e_{\nu}^{j}\omega^{i}{}_{j\mu}+e_{i}^{\lambda}\delta e_{\nu}^{j}\omega^{i}{}_{j\mu}+e_{i}^{\lambda}e_{\nu}^{j}\delta\omega^{i}{}_{j\mu}+\delta e_{i}^{\lambda}\partial_{\mu}e_{\nu}^{i}\right)\right.\\
 & \phantom{{}=}\left.-\delta e_{\nu}^{i}\left(\nabla_{\mu}\mathfrak{\mathfrak{N}}_{i}{}^{\mu\nu}+\omega^{j}{}_{i\mu}\mathfrak{\mathfrak{N}}_{j}{}^{\mu\nu}-\Gamma^{\mu}{}_{\rho\mu}\mathfrak{\mathfrak{N}}_{i}{}^{\rho\nu}-\Gamma^{\nu}{}_{\rho\mu}\mathfrak{\mathfrak{N}}_{i}{}^{\mu\rho}\right)\right)\mathrm{d}^{4}x\\
 & =2\int\left(e_{i}^{\lambda}e_{\nu}^{j}\mathfrak{\mathfrak{N}}_{\lambda}{}^{\mu\nu}\delta\omega^{i}{}_{j\mu}-\left(\nabla_{\mu}\mathfrak{\mathfrak{N}}_{i}{}^{\mu\nu}-\Gamma^{\mu}{}_{\rho\mu}\mathfrak{\mathfrak{N}}_{i}{}^{\rho\nu}\right)\delta e_{\nu}^{i}\right.\\
 & \phantom{{}=}+e_{i}^{\lambda}\omega^{i}{}_{j\mu}\mathfrak{\mathfrak{N}}_{\lambda}{}^{\mu\nu}\delta e_{\nu}^{j}-\omega^{j}{}_{i\mu}\mathfrak{\mathfrak{N}}_{j}{}^{\mu\nu}\delta e_{\nu}^{i}\\
 & \phantom{{}=}+\mathfrak{\mathfrak{N}}_{\lambda}{}^{\mu\nu}\left(e_{\nu}^{j}\omega^{i}{}_{j\mu}\delta e_{i}^{\lambda}+\partial_{\mu}e_{\nu}^{i}\delta e_{i}^{\lambda}\right)\\
 & \phantom{{}=}\left.+\mathfrak{\mathfrak{N}}_{i}{}^{\mu\rho}\Gamma^{\nu}{}_{\rho\mu}\delta e_{\nu}^{i}\right)\mathrm{d}^{4}x\\
 & =2\int\left(-\mathfrak{\mathfrak{N}}_{i}{}^{j\mu}\delta\omega^{i}{}_{j\mu}-\left(\nabla_{\mu}\mathfrak{\mathfrak{N}}_{i}{}^{\mu\nu}-\Gamma^{\rho}{}_{\mu\rho}\mathfrak{\mathfrak{N}}_{i}{}^{\mu\nu}\right)\delta e_{\nu}^{i}\right.\\
 & \phantom{{}=}+\left(\omega^{i}{}_{j\mu}\mathfrak{\mathfrak{N}}_{i}{}^{\mu\nu}-\omega^{i}{}_{j\mu}\mathfrak{\mathfrak{N}}_{i}{}^{\mu\nu}\right)\delta e_{\nu}^{j}\\
 & \phantom{{}=}+\frac{1}{2}\mathfrak{\mathfrak{N}}_{\lambda}{}^{\mu\nu}\left(e_{\nu}^{j}\omega^{i}{}_{j\mu}-e_{\mu}^{j}\omega^{i}{}_{j\nu}+\partial_{\mu}e_{\nu}^{i}-\partial_{\nu}e_{\mu}^{i}\right)\left(-e_{i}^{\rho}e_{k}^{\lambda}\delta e_{\rho}^{k}\right)\\
 & \phantom{{}=}\left.+\frac{1}{2}\mathfrak{\mathfrak{N}}_{i}{}^{\mu\nu}\left(\Gamma^{\lambda}{}_{\nu\mu}-\Gamma^{\lambda}{}_{\mu\nu}\right)\delta e_{\lambda}^{i}\right)\mathrm{d}^{4}x\\
 & =2\int\left(-\mathfrak{\mathfrak{N}}_{i}{}^{j\mu}\delta\omega^{i}{}_{j\mu}-\left(\nabla_{\mu}+T^{\rho}{}_{\mu\rho}\right)\mathfrak{\mathfrak{N}}_{i}{}^{\mu\nu}\delta e_{\nu}^{i}\right.\\
 & \phantom{{}=}-\frac{1}{2}\mathfrak{\mathfrak{N}}_{\lambda}{}^{\mu\nu}T^{\rho}{}_{\mu\nu}e_{k}^{\lambda}\delta e_{\rho}^{k}\\
 & \phantom{{}=}\left.+\frac{1}{2}\mathfrak{\mathfrak{N}}_{i}{}^{\mu\nu}T^{\lambda}{}_{\mu\nu}\delta e_{\lambda}^{i}\right)\mathrm{d}^{4}x\\
 & =2\int\left(-\mathfrak{\mathfrak{N}}_{i}{}^{j\mu}\delta\omega^{i}{}_{j\mu}+\left(\nabla_{\lambda}+T^{\rho}{}_{\lambda\rho}\right)\mathfrak{\mathfrak{N}}_{i}{}^{\mu\lambda}\delta e_{\mu}^{i}\right)\mathrm{d}^{4}x,
\end{aligned}
\end{equation}
where we use the anti-symmetry of $\mathfrak{N}$ and relabel dummy
indices throughout, and in the fourth line we use the divergence theorem
recalling that $\delta e_{\nu}^{i}$ vanishes on the boundary to find
that
\begin{equation}
\begin{aligned}\int\mathfrak{\mathfrak{N}}_{\lambda}{}^{\mu\nu}e_{i}^{\lambda}\partial_{\mu}\left(\delta e_{\nu}^{i}\right)\mathrm{d}^{4}x & =\int\partial_{\mu}\left(\mathfrak{\mathfrak{N}}_{\lambda}{}^{\mu\nu}e_{i}^{\lambda}\delta e_{\nu}^{i}\right)\mathrm{d}^{4}x-\int\delta e_{\nu}^{i}\partial_{\mu}\left(\mathfrak{\mathfrak{N}}_{\lambda}{}^{\mu\nu}e_{i}^{\lambda}\right)\mathrm{d}^{4}x\\
 & =-\int\delta e_{\nu}^{i}\partial_{\mu}\mathfrak{\mathfrak{N}}_{i}{}^{\mu\nu}\mathrm{d}^{4}x,
\end{aligned}
\end{equation}
and then express the partial derivative in terms of the covariant
one.

Equating the terms in $\delta\omega$ after anti-symmetrizing yields
\begin{equation}
\begin{aligned}\frac{1}{2}\int\mathfrak{S}_{i}{}^{j\mu}\delta\omega^{i}{}_{j\mu}\mathrm{d}^{4}x & =-2\int\frac{1}{2}\left(\mathfrak{\mathfrak{N}}_{i}{}^{j\mu}-\mathfrak{\mathfrak{N}}_{j}{}^{i\mu}\right)\delta\omega^{i}{}_{j\mu}\mathrm{d}^{4}x\\
 & =-\frac{1}{4}\int\left(\mathfrak{S}^{ji\mu}+\mathfrak{S}^{j\mu i}-\mathfrak{S}^{\mu ij}-\mathfrak{S}^{ij\mu}-\mathfrak{S}^{i\mu j}+\mathfrak{S}^{\mu ji}\right)\delta\omega_{ij\mu}\mathrm{d}^{4}x\\
 & =\frac{1}{2}\int\mathfrak{S}^{ij\mu}\delta\omega_{ij\mu}\mathrm{d}^{4}x
\end{aligned}
\end{equation}
as expected, using the anti-symmetry of the first two indices in $\mathfrak{S}$.
Equating the terms in $\delta e$ yields
\begin{equation}
-\begin{aligned}\int\overset{+}{\mathfrak{T}}_{\mu}{}^{i}\delta e_{i}^{\mu}\mathrm{d}^{4}x & =-\int\overline{\mathfrak{T}}_{\mu}{}^{i}\delta e_{i}^{\mu}\mathrm{d}^{4}x-2\int\left(\nabla_{\lambda}+T^{\rho}{}_{\lambda\rho}\right)\mathfrak{\mathfrak{N}}_{i}{}^{\mu\lambda}\delta e_{\mu}^{i}\mathrm{d}^{4}x.\end{aligned}
\end{equation}
In order to match variations we calculate that 
\begin{equation}
\begin{aligned}\mathfrak{\mathfrak{N}}_{i}{}^{\mu\lambda}\delta e_{\mu}^{i} & =-\mathfrak{\mathfrak{N}}_{i}{}^{\mu\lambda}e_{\nu}^{i}e_{\mu}^{j}\delta e_{j}^{\nu}\\
 & =-\mathfrak{\mathfrak{N}}_{\nu}{}^{j\lambda}\delta e_{j}^{\nu}\\
 & =-\mathfrak{\mathfrak{N}}_{\mu}{}^{i\lambda}\delta e_{i}^{\mu}\\
 & =\mathfrak{\mathfrak{N}}_{\mu}{}^{\lambda i}\delta e_{i}^{\mu}
\end{aligned}
\end{equation}
and recall that the covariant derivative of both the metric and tetrad
vanishes, yielding the Belinfante-Rosenfeld relation
\begin{equation}
\begin{aligned}-\overset{+}{\mathfrak{T}}_{\mu}{}^{i} & =-\overline{\mathfrak{T}}_{\mu}{}^{i}-2\left(\nabla_{\lambda}+T^{\rho}{}_{\lambda\rho}\right)\mathfrak{\mathfrak{N}}_{\mu}{}^{\lambda i}\\
\Rightarrow\overline{\mathfrak{T}}{}^{\mu\nu} & =\overset{+}{\mathfrak{T}}{}^{\mu\nu}-2\left(\nabla_{\lambda}+T^{\rho}{}_{\lambda\rho}\right)\mathfrak{\mathfrak{N}^{\mu\lambda\nu}}\\
\Rightarrow\overline{T}{}^{\mu\nu} & =\overset{+}{T}{}^{\mu\nu}-\frac{1}{2}\left(\nabla_{\lambda}+T^{\rho}{}_{\lambda\rho}\right)\left(S^{\lambda\mu\nu}+S^{\lambda\nu\mu}-S^{\nu\mu\lambda}\right).
\end{aligned}
\label{eq:belinfante-rosenfeld-relation-alt-deriv}
\end{equation}

Note that unlike our previous one, this derivation does not use the
gravitational action, and therefore is valid for any such action.
In particular, we may consider a gravitational Lagrangian which consists
of the torsionless scalar curvature $\overline{R}$, despite the matter
action depending upon a covariant derivative $\nabla$ which includes
torsion. For such a gravitational action we have $\mathring{T}=0$,
so that the EOM are $\overline{G}=\kappa\overline{T}$ and $S=0$,
but the Belinfante-Rosenfeld relation nevertheless holds for the matter
action; when the EOM are satisfied, we therefore have $\overline{G}=\kappa\overline{T}=\kappa\overset{+}{T}$,
and hence the tetrad SEM tensor will be symmetric and divergenceless.

\end{document}